\documentclass[aps,prd,letterpaper,twocolumn,superscriptaddress,floatfix]{revtex4-2}

\usepackage{graphicx,psfrag,amsmath,amssymb,amsfonts,bbm,latexsym,color,xcolor,dcolumn,bm,braket,times}
\usepackage[english]{babel}

\usepackage[
    starfontserif % comment for sans glyphs
    ]{starfont}
\DeclareSymbolFont{starfontsym}{OT1}{sts}{m}{n}
\DeclareMathSymbol{\mathSun}{\mathord}{starfontsym}{115}
\DeclareMathSymbol{\mathTerra}{\mathord}{starfontsym}{76}

\usepackage{soul}
\newcommand{\nn}{\nonumber}

\usepackage{colonequals} 				
\newcommand{\ce}{\colonequals}

\begin{document}
	
\title{Testing the weak equivalence principle for nonclassical matter with torsion balances}

\author{Roberto Onofrio}
\affiliation{\mbox{Department of Physics and Astronomy, Dartmouth College, 6127 Wilder Laboratory, Hanover, NH 03755, USA}}
		
\author{Alexander R. H. Smith}
\affiliation{\mbox{Department of Physics, Saint Anselm College, Manchester, NH 03102, USA}}
\affiliation{\mbox{Department of Physics and Astronomy, Dartmouth College, 6127 Wilder Laboratory, Hanover, NH 03755, USA}}
		
\author{Lorenza Viola}
\affiliation{\mbox{Department of Physics and Astronomy, Dartmouth College, 6127 Wilder Laboratory, Hanover, NH 03755, USA}}

\date{\today}
 
\begin{abstract}
We propose tests of the weak equivalence principle (WEP) using a torsion balance, in which superposition of energy eigenstates are created in a controllable way for the test masses. After general considerations on the significance of tests of the WEP using quantum states and the need for considering inertial and gravitational masses as operators, we develop a  model to derive the matrix elements of the free-fall operator, showing that the variance of the acceleration operator, in addition to its mean, enables estimation of violations of the WEP due to quantum coherence in a way that is robust with respect to shot-to-shot fluctuations. Building on this analysis, we demonstrate how the validity of the WEP may be tested in a torsion balance setup, by accessing the mean and variance of a torque operator we introduce and quantize. Due to the long acquisition times of the signal as compared to the timescale on which coherent superposition states may survive, we further propose a dynamical setting, where the torsion balance is subject to a time-dependent gravitational field, and measurements of angular acceleration encode possible violations of the WEP.  
\end{abstract}

\maketitle

\section{Introduction}

In its original classical formulation, the weak equivalence principle (WEP) states that~\cite{Will1993,Will2014}:
 If an uncharged test particle is placed at an initial event in spacetime and given an initial velocity 
 there, then its subsequent trajectory will be independent of its internal structure and composition.
As a consequence, the WEP bounds the relative acceleration in free fall between two pointlike test masses\,---\,say,  
$A$ and $B$, each falling with acceleration ${\bf a}_A$ and ${\bf a}_B$ in the presence of a fixed external gravitational field\,---\,to be zero; equivalently,  ${\bf a}_A={\bf a}_B$ for any choice of the test masses. 
This principle, more than any other, has played an important role in the development of gravitational physics beginning with Newton's {\em Principia}, 
it has played a heuristic role in the conception of any metric theory of gravity such as general relativity, and has been crucial in ruling out 
possible variants of gravitational theories in which the coupling of the energy-momentum tensor to material-specific properties occurs. 

Within Newtonian gravity, the acceleration of a body is related to its inertial mass, whereas the 
gravitational force is related to the gravitational ``charge,'' also referred to as the gravitational mass. 
The WEP then demands the equality of inertial and gravitational mass: if this were not the case, the acceleration, 
and thus the subsequent trajectory, of an object falling in a gravitational field would depend upon its internal structure. 
In general relativity, the WEP still implies universality of free fall, in that free particles follow geodesic trajectories through spacetime, independently of their composition and internal structure.
 
Conceptually\,---\,and in line with the ideal ``leaning-tower'' experiment discussed by Galileo \cite{Galileo}\,---\,the simplest way to test the WEP is to use bodies in free fall in the Earth's gravitational field. In this setting, the violation of the WEP is quantified by the {\em E\"{o}tv\"{o}s ratio}~\cite{Will1993}, defined as 
$\eta \ce 2 |{\bf a}_A - {\bf a}_B|/|{\bf a}_A + {\bf a}_B|$, where ${\bf a}_A$ and ${\bf a}_B$ are the acceleration of two freely falling bodies $A$ and $B$. 
However, the achievable sensitivity to violations is limited, among other sources, by the uncertainty in controlling the 
initial conditions and determining the duration of the fall. This has prompted the use of differential free-fall measurements, 
in which the {\em relative} motion of two falling bodies is monitored, and a null result is targeted~\cite{Polacco1,Polacco2}. 
In this context, the WEP has been tested in tabletop experiments using self-adjusting torsion balances in which torques are simultaneously produced by gravity and the noninertial motion of the balance due to the Earth's rotation\,---\,the former being 
proportional to the gravitational mass and the latter to the inertial mass. 
The torque $\tau_{\parallel}$ about the balance is then proportional to the E\"{o}tv\"{o}s ratio.
In this way, a nonzero value of $\tau_{\parallel}$ implies a violation of the WEP.
Experimental bounds on the E\"{o}tv\"{o}s ratio obtained from torsion balance experiments have a precision on the order of 
$\eta \lesssim 10^{-13}$ in tabletop experiments  \cite{Schlamminger2008,Adelberger2009,Zhu2018} 
and $\eta \lesssim 10^{-15}$ using macroscopic masses in artificial satellites~\cite{Touboul2019a,Touboul2019b,Touboul2022}. 

Considering the crucial role of the WEP, the importance of testing its validity in a domain where the test masses and their dynamics must be described using quantum mechanics has long been appreciated \cite{Lammerzahl1996, Viola1997}. 
Violations of the WEP could indicate a possible relationship between gravity and the other fundamental interactions in nature, all of which are flavor dependent, unlike the gravitational interaction. However, it is not {\em a priori} obvious how the WEP ought to be applied to a quantum system, for which the notion of a spacetime trajectory may be problematic~\cite{Sonego1995,Anastopoulos2018}.
Indeed, in the nonrelativistic weak-field limit, mass-dependent uncertainties are predicted for time-of-flight measurements 
of generic superposition states of {\em external} translational degrees of freedom in free-fall experiments of atoms~\cite{Viola1997}. This calls for care in interpreting and applying the WEP in the quantum regime, including more recent precision tests using interferometry of cold atoms \cite{Peters1999,Fray2004,Schlippert2014}.

Crucially, in addition to allowing for more general initial preparations of ``structureless'' test masses, quantum mechanics grants the possibility of qualitatively different {\em internal} structures, arising from coherent superposition between quantum states and entanglement among constituent degrees of freedom~\cite{Zych2018,Zych2019}. 
It then becomes both natural and important to determine the validity of the WEP in regard to such compositional structures. 
A major experimental advance in this direction was achieved by Rosi {\it et al.} by using an atomic interferometer to bound violations of the WEP due to internal quantum coherence \cite{Rosi2017}. The relevant quantity in this case is the operator $\hat{M}_{g} \hat{M}_{i}^{-1}$, where $\hat{M}_{g}$ and $\hat{M}_{i}$ are gravitational and inertial mass operators, respectively, for a test body in free fall. These mass operators are proportional to the internal energy of the test body through mass-energy equivalence and, in the simplest case, $\hat{M}_{g} \hat{M}_{i}^{-1}$ admits a $2 \times 2$ matrix representation when the addressable internal-energy states correspond to a two-level system. The diagonal matrix elements of $\hat{M}_{g} \hat{M}_{i}^{-1}$  must be equal to unity if the classical WEP holds; however, the possibility arises for $\hat{M}_{g} \hat{M}_{i}^{-1}$ to have an {\em off-diagonal} matrix element, expressing a genuine violation of the WEP without a classical counterpart. The experiment resulted in the first bound being placed on the modulus of the off-diagonal matrix element $|r|$ of $\hat{M}_{g} \hat{M}_{i}^{-1}$, constraining it to be $|r| \lesssim 10^{-8}$~\cite{Rosi2017}. The bounds on the {\em diagonal} matrix elements were stronger by one order of magnitude, and have been superseded by a more recent experiment \cite{Asenbaum2020}, which also used interferometric techniques and resulted in a bound  of $10^{-12}$. These bounds are still not competitive with the ones from table-top torsion balances or satellites quoted above, that test the WEP in the classical regime. A careful discussion of the systematic errors preventing atom interferometers from reaching higher sensitivities is presented in \cite{Nobili2016}, with emphasis on the degree of control of the initial conditions for average position and momentum of cold atomic clouds (see also \cite{Pfeiffer} for a recent analysis). 

Given the  success of torsion balances in bounding violations of the WEP in the classical regime, our main aim is to 
propose torsion balance experiments that open up new possibilities for precision tests of the validity of the WEP in a quantum regime. We note that our approach is qualitatively different from existing proposals that have been put forward for quantum extensions of the WEP
based on properties of the center-of-mass wave-functions of falling particles~\cite{Anastopoulos2018}, the existence of local quantum reference frames exhibiting definite causal order in superpositions of spacetime~\cite{Hardy2018,Hardy2019}, 
or even superpositions of quantum reference frames~\cite{Giacomini2022,Giacomini2024}.

The content of the paper is organized as follows. In Sec.\,II, we revisit the motivations for promoting the inertial and 
gravitational masses to quantum operators, and some of the consequences, especially regarding theories of gravitation.
We also review a context in which meaningful questions on the existence or absence of genuine quantum violations of the 
WEP can be asked and answered. This allows us to focus on an ideal experiment in genuine free fall to emphasize how 
in principle the entire set of moments of the acceleration observable is crucial to test the WEP.  
The analysis shows that the average acceleration and its variance are robust indicators in the presence of uncontrollable 
phase factors that are inevitable in the initial internal state of the falling test body, whose values vary randomly for different realizations of the experiment.
In Sec.\,III, we examine an E\"{o}tv\"{o}s-like experiment in which the internal energy of the torsion balance test masses can be placed into a quantum superposition. Such considerations necessitate promoting the torques that act on the torsion balance to quantum operators themselves. We evaluate the average torque on the balance and its variance, demonstrating that these moments are sensitive to quantum violations of the WEP due to internal quantum coherence of the balance's test masses.
In Sec.\,IV, we consider a more realistic dynamical Cavendish experiment in which the test masses of a torsion balance are again placed in a quantum superposition of internal energy levels, but the gravitational force is instead provided by local masses rotating about the balance, as opposed to the the gravitational force of the Earth or Sun in an E\"{o}tv\"{o}s-like experiment.
We discuss the possibility of a WEP test with torsion balances in the light of state-of-art capabilities in 
producing massive quantum superposition states, as well as in preserving quantum coherence through quantum control methods.  
Two factors are crucial to perform tests of the WEP involving quantum matter and torsional balances. First, the possibility to create superpositions of quantum states for a significant percentage of atoms, exploiting the continuous and fast progress in preparing qubits in solid-state devices. Second, the availability of sources of dynamical gravitational fields used as calibrators for gravitational wave antennas, which enables modulation of the gravitational signal at frequencies comparable (or higher) than the decoherence rate of the quantum superposition. While the proposed experiments are not yet within reach, we hope that our discussion will stimulate and provide further motivation for this crucial subfield of quantum metrology. More general considerations complete the paper, with Appendices complementing the discussions in Secs.\,III and IV.

\section{Gravitational and inertial masses\\ as operators and the  WEP}

\label{Gravitational and inertial masses as operators}

In this Section, we first discuss the motivations for introducing gravitational and inertial mass as operators and related issues. 
We then describe a simple model in which such operators can be discussed quantitatively. 

\subsection{Mass as an operator}

In discussing the  WEP in the quantum realm, it is necessary to consider test bodies 
whose states, dynamics, and measurement are described to the full level of generality permitted by quantum-mechanical laws. 
Treating test bodies as approximately classical is possible only in the limit of large masses; conceptually, however, this conflicts 
with the very notion of a test body, as in such a setting the assumption of a minimal perturbation to the preexisting gravitational 
fields becomes questionable. Besides this operational issue, a longstanding debate exists on the need to treat the inertial mass as an immutable, scalar parameter of a system or as an operator. Considering the formulation of the WEP in classical mechanics, this 
has also implications on the concept of gravitational mass. The issue is far from being settled from both a conceptual and operational standpoint, as we summarize in the following. 

On the one hand, the standard model of elementary particle physics supports the view that the inertial mass arises from 
the vacuum expectation value of the Higgs field operator via (for now) {\em ad hoc} Yukawa coupling coefficients. 
Modifying the vacuum expectation value of the Higgs field, one will presumably obtain (if the Yukawa couplings are
fixed parameters) different values for the mass of elementary particles. If so, the inertial mass has to be considered 
as an operator, also implying the possibility of fluctuations around its average value in an assigned state\,---\,although those would presumably occur on a timescale $\Delta t \simeq \lambda_H/c$ settled by the Compton wavelength 
of the Higgs field, $\lambda_H= h/(M_H c)$, with $M_H$ being the Higgs mass. Ideally, the presence of well-defined values of 
mass for elementary particles could be justified by a future unified scheme in which each inertial mass is obtained as an 
eigenvalue of a generalized mass operator. Independent considerations, unrelated to the presence of the Higgs field, and instead 
based on the relativity of quantum vacuum, also point to the possibility of inertial mass being an operator \cite{Jaekel1997}.

On the other hand, considering the inertial mass as an operator raises a number of issues, as it implies the possibility of superposition of 
mass eigenstates, which (thus far) we have no experimental evidence for, and which complicates the dynamics already at the 
nonrelativistic level \cite{Giulini}, due to the presence of the mass in the kinetic energy term of the Schr\"{o}dinger equation. 
The problem may be circumvented, accommodating the presumed lack of evidence for mass superposition states, by postulating the impossibility of creating such states\,---\,via a superselection rule \cite{Bargmann1954}. 
This adds an extra feature to the postulates of quantum mechanics, undermining the foundational economicity 
of the latter. Attempts to remove the superselection rule tend to complicate the dynamics, for instance through the 
introduction of another parameter that assumes the meaning of an ``internal'' time for the system \cite{Coronado2013} 
(see, however, a recent discussion in \cite{ZychGreen}). All of the above points to the convenience of considering
 the mass as a scalar parameter.

Notwithstanding the contrasting arguments above, we can safely state that the ``internal'' energy of a body, due to interactions 
among its components, necessitates an operational description as customary for bound states in atoms and molecules \cite{Lammerzahl1995}. 
In the spirit of the extensive discussion in \cite{Zych2018}, here we assume the operator nature of inertial mass as a {\em working hypothesis}, possibly to be tested against all possible consequences at the dynamical level. Furthermore, to meaningfully discuss a quantum extension of the WEP, we also consider the gravitational mass as an operator, for otherwise conceptual inconsistencies arise at the outset.
With this in mind, we can imagine that the inertial mass of a test body suitable to study the WEP is comprised of a classical contribution 
$m_i$ and a 
``quantum'' contribution stemming from the internal energy $\hat{H}_{i}/c^2$, the sum of which define the so-called {\em inertial mass operator}, $\hat{M}_{i} = m_{i} + \hat{H}_{i}/c^2$, where $m_{i}$ is meant to be multiplied by an identity operator $\hat{I}$ of the appropriate dimension. 
Associated with the test body is also a gravitational mass operator $\hat{M}_{g} = m_{g} + \hat{H}_{g}/c^2$, which {\em a priori} may or may not be equal to the inertial mass operator $\hat{M}_{i}$. As discussed in~\cite{Rosi2017,zych_quantum_2017,Zych2018,Zych2019, Das2023}, a natural formulation of the WEP for quantum systems is then to demand equality of the inertial and gravitational mass operators, $\hat{M}_{i} =\hat{M}_{g}$.

\subsection{Approximate free-fall acceleration operator}
\label{sub:approx}

Building on the above considerations, we now discuss the simplest setting that captures the operator nature of the 
mass, by focusing on quantum systems in which two internal energy levels are addressable at the microscopic level, and 
discuss the consequences for free-fall experiments. 

The E\"{o}tv\"{o}s ratio for two classical test bodies $A$ and $B$ can be expressed in terms of the inertial and gravitational masses $M_{i,j}$ and $M_{g,j}$, respectively, with $j \in \{A,B\}$ \cite{Will1993,Will2014}, as
\begin{equation}
\eta =2 \frac{\left| M_{g,A} M_{i,A}^{-1} - M_{g,B} M_{i,B}^{-1}\right|}{ \left|M_{g,A} M_{i,A}^{-1} + M_{g,B} M_{i,B}^{-1} \right|}.
\label{Eotvos}
\end{equation}
In a free-fall experiment aimed at evidencing the role of the WEP for the internal energy, there is no gain in 
using two masses of different composition; therefore, we can imagine that $A$ and $B$ correspond to two distinct initial preparations 
of the same test mass. This choice is even more adequate in the setting of free-fall experiments using atomic interferometry, as in this case the test masses are typically atoms of the same species~\cite{Rosi2017}. Even in atom-interferometry tests that use different species, as in the pioneering experiment reported in~\cite{Fray2004}, each run involves only one initial preparation of the test system, so the above interpretation still applies for each species.

Our goal is to extend the E\"{o}tv\"{o}s parameter to the quantum regime, assuming that both gravitational and inertial masses are described by operators. Before tackling a concrete experimental setup, we discuss the form of the mass operators for a two-level quantum system, defined on a Hilbert space $\mathcal{H} \simeq \mathbb{C}^2$. One should keep in mind that neither a torsion balance nor atoms in an atomic interferometer are actually in free fall, although the latter spend only a minute fraction (order of 0.1$\,\%$) of their interrogation time interacting with lasers.

Specifically, we are interested in different compositions of internal mass and gravitational mass {\em arising from different amounts of quantum coherence between internal energy levels}. For this reason, we will also suppose that the center-of-mass degree of freedom of both $A$ and $B$ is well localized and follows a classical trajectory. 

In this spirit, our first step is to write an approximate form for the operator $\hat{M}_g \hat{M}_i^{-1}$ in terms of the matrix elements of the internal inertial and gravitational Hamiltonians, $\hat{H}_i$ and $\hat{H}_g$, which in realistic situations are expected to be very small with respect to the non-dynamical inertial and gravitational masses, $m_i$ and $m_g$. Following Ref.\,\cite{Rosi2017} [see Eq.\,(3) therein], recall that the free-fall acceleration operator may be expressed as follows with respect to the eigenbasis of $\hat{H}_i$,
\begin{equation}
\hat{a}= g \hat{M}_g \hat{M}_i^{-1} \ce g 
\begin{pmatrix}
r_1  & r \\
r^* &  r_2
\end{pmatrix},
\label{aFreeFall}
\end{equation}
where  $r_1, r_2 \in {\mathbb R}$ and the complex parameter $r\ce |r| e^{i \varphi_r}$, with $\varphi_r \in [0, 2\pi)$, and $g$ the local gravitational acceleration. A necessary and sufficient condition for the WEP to be valid is that $r_1 = r_2 =1$ and $r =0$, so that $\hat{a}=g\hat{I}$. 

We work to the leading order in $\hat{H}_i/m_ic^2$ and $\hat{H}_g/m_ic^2$, in which case $[\hat{M}_g,\hat{M}_i] = 0$, hence the ordering 
$\hat{M}_g$ and $\hat{M}_i^{-1}$ in the definition of $\hat{a}$ is unimportant. In this approximation, the free-fall acceleration operator is 
\begin{align}
    \hat{a} &\approx g \frac{m_g}{m_i}\bigg( \hat{I} + 
    \frac{\hat{H}_g}{m_gc^2}-
    \frac{\hat{H}_i}{m_ic^2}\bigg) \nonumber \\
    &=  g \frac{m_g}{m_i} 
\begin{pmatrix}
1+\frac{E_{11}^{(g)}}{m_g c^2}-\frac{E_{11}^{(i)}}{m_i c^2}  & \frac{E_{12}^{(g)}}{m_g c^2} \\
\frac{E_{12}^{(g)*}}{m_g c^2} &  1+\frac{E_{22}^{(g)}}{m_g c^2}-\frac{E_{22}^{(i)}}{m_i c^2}
\end{pmatrix},
\label{aApprox}
\end{align}
where $E_{jk}^{(g)}$ and $E_{jk}^{(i)}$ are, respectively, the matrix elements of $\hat{H}_{g}$ and $\hat{H}_{i}$ 
in the eigenbasis $\{|1\rangle, |2\rangle\}$ of $\hat{H}_i$. It follows that 
\begin{align}
    r_1 &= \frac{m_g}{m_i} \left[ 1+E_{11}^{(g)}/(m_g c^2)-E_{11}^{(i)}/(m_i c^2) +  \dots \right], \label{r1} \\
    r_2 &= \frac{m_g}{m_i} \left[ 1+E_{22}^{(g)}/(m_g c^2)-E_{22}^{(i)}/(m_i c^2) +  \dots\right], \label{r2} \\
    r &=\frac{m_g}{m_i} \left[  E_{12}^{(g)}/(m_g c^2) +  \dots \right].
    \label{rdiag}
\end{align}

Suppose that the internal state of the particles making up the test mass in free fall is described by a general density operator on $\mathbb{C}^2$, of the form
\begin{equation}
    \hat{\rho}(n,\theta,\phi) 
    = \frac{1}{2} \left(\hat{I} + \vec{n} \cdot \vec{\sigma} \right) , 
    \label{TwoLevelState}
\end{equation}
where $\vec{\sigma} \ce (\sigma_x,\sigma_y,\sigma_z)$ is a vector of Pauli matrices and $\vec{n}\in {\mathbb R}^3$ is the Bloch 
(or coherence) vector, $\vec{n}\ce (n_x,n_y,n_z)= ( \langle \sigma_x \rangle,  \langle\sigma_y\rangle, \langle\sigma_z\rangle )$. 
In spherical coordinates, we let $\vec{n} \ce n\, (\sin \theta \cos \phi, \sin \theta \sin \phi, \cos \theta)$, in terms of the two angles 
$\theta \in [0, \pi]$ and $\phi\in[-\pi, \pi)$, with $n \leq 1$ due to the non-negativity of $\hat{\rho}$. 
For pure states, $n=1$ and $\hat{\rho}(n,\theta,\phi) =|\psi(\theta, \phi)\rangle \langle \psi(\theta, \phi)|$, with $\theta$ and $\phi$ identifying 
the direction of the state-vector on the Bloch  sphere. For a general state,  a measure of coherence in the energy basis is provided by \cite{Plenio}
\begin{equation}
\sum_{j\ne k}|\rho_{jk}(n,\theta,\phi)| = \big({n_x^2 +n_y^2}\big)^{1/2}=  n\sin\theta , \nn
\end{equation}
showing how coherence is maximized for $\theta =\pi/2$ at fixed $n$. By assuming that the measurement of acceleration takes place on a timescale much shorter relative to the timescale $\sim\!h/\Delta E_i$ associated to the internal evolution under $\hat{H}_i$, this evolution may be neglected. Given the above parametrization, the mean value of the free-fall acceleration operator, normalized to $g$, can then be obtained as
\begin{align}
\!\!\frac{{\langle\hat{a}\rangle}_{\rho}}{g} &= F(r_1,r_2,|r|,\varphi_r ; n,\theta,\phi) \ce \frac{r_1 + r_2}{2}  
\nonumber \\ 
& \quad +  n \bigg[ \frac{r_1-r_2}{2} \cos \theta +  |r| \cos \left(\varphi_r + \phi \right) \sin \theta \bigg],
\label{Average} 
\end{align}
while its variance, $(\Delta \hat{a}^2)_\rho= \langle \hat{a}^2 \rangle_\rho - \langle \hat{a} \rangle_\rho^2$, reads
\begin{align}
\!\!\!\frac{(\Delta \hat{a}^2)_\rho}{g^2} & = G(r_1,r_2,|r|,\varphi_r ; n,\theta,\phi) 
\ce \bigg( \frac{r_1 - r_2}{2} \bigg)^{\!2} + |r|^2  \nonumber \\
& \quad  -n^2 \left[ \frac{r_1-r_2}{2} \cos \theta + |r| \cos\left( \varphi_r + \phi \right) \sin \theta \right]^2 \!. 
\label{AAverage}
\end{align} 
In the above expressions, we have introduced two form factors, $F$ and $G$, which depend upon all the kinematic 
and dynamical variables that parametrize possible violations of the WEP, {\it i.e.,} the four parameters ($r_1,r_2,r,\varphi_r$) 
appearing in Eq.\,\eqref{aFreeFall} for a complete description of the free-fall acceleration operator, as well as the 
coherence-vector components $(n,\theta,\phi)$ describing the internal state of the test mass in Eq.\,\eqref{TwoLevelState}. 

\begin{figure*}
\centering
\qquad\includegraphics[width=16cm]{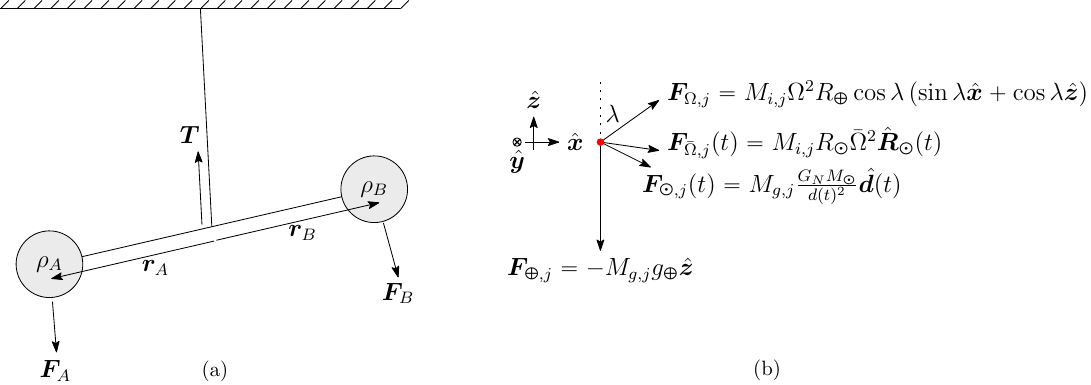}
%\qquad\includegraphics[width=0.9\textwidth]{Fig1.pdf}
\vspace*{-2mm}
\caption{(a) Torsion balance in which each arm comprises a collection of two-level quantum systems (e.g., a solid-state high-density NV ensemble) whose state, $\hat{\rho}_A$ and $\hat{\rho}_B$, can be coherently controlled and placed into a superposition of internal energy states. ${\bf F}_A$ and ${\bf F}_B$ are the net forces acting on each arm at distances ${\bf r}_A$ and ${\bf r}_B$ from the center of the balance and ${\bf T}$ is the tension in the fiber. (b) Depicted are the forces acting on one arm of the balance, labeled by $j \in \{A,B\}$, in the lab-centered, noninertial frame. ${\bf F}_{C,j}$ is the centrifugal force, which is proportional to the inertial mass $M_{i,j}$, while ${\bf F}_{\mathSun,j}$ and ${\bf F}_{\mathTerra,j}$ are the gravitational force of the Sun and Earth, respectively, which are proportional to the gravitational mass $M_{g,j}$; the $z$ axis is chosen to align with the local gravitational field of the Earth. Not taken into account are the gravitational force of the Moon, the centrifugal force due to the orbital motion of the Earth around the Sun, and perturbations induced by other Solar System objects.}
\label{Fig:eotvos}
\end{figure*}

While both the first and second cumulants of the acceleration of the falling test body are seen to depend sensitively on the internal state
and the validity of the WEP, Eq.\,\eqref{AAverage} differs from Eq.\,\eqref{Average} in an important way: namely, in the function $G$, 
a term that depends {\em solely upon $|r|$} is present, which is advantageous for separately assessing diagonal and off-diagonal contributions in $\hat{a}$.
More specifically, there is a dependence upon $|r|$ even in the case where the internal state is completely incoherent (maximally mixed, $n=0$); a combined analysis of the mean value and the variance of the acceleration allows, by subtraction, to assess bounds (or a meaningful nonzero value) to $|r|$. If instead the state is only partially mixed or pure ($ 0< n \leq 1 $), control over its preparation makes it possible to disentangle the two contributions, proportional to $r_1-r_2$ and $|r|$. This may be clearer upon rearranging the 
terms in the right-hand side of Eq.\,\eqref{AAverage} to get
\begin{align}
\frac{(\Delta \hat{a}^2)_\rho}{g^2}&= \bigg(\frac{r_1 - r_2}{2}\! \bigg)^{\!2}\! \Big ( 1- n^2 \cos^2 \theta \Big)\notag \\   
 &\quad +  |r|^2 \Big[ 1- n^2 \cos^2(\varphi_r +\phi) \sin^2 \theta \Big] \label{Apure} \\
& \quad  - 2\, n^2 |r| \,\bigg(\frac{r_1-r_2}{2} \!\bigg)\cos(\varphi_r+\phi) \sin \theta \cos\theta .\nonumber
\end{align}
For the two pure states $|\psi_1\rangle \ce|1\rangle$ $(n=1; \theta=\phi=0)$ and $|\psi_+\rangle \ce \tfrac{1}{\sqrt{2}} (|1\rangle +|2\rangle)$ $(n=1; \theta=\pi/2, \phi=0)$, this equation yields the values of $|r|^2$ and $(r_1-r_2)^2/4+|r|^2 \sin^2(\varphi_r)$, respectively. In situations where the relative phase $\phi$ between inertial energy eigenstates is uncontrollable, as in \cite{Rosi2017}, the resulting shot-to-shot variation translates into phase randomization when averaging over multiple realizations. To compare with theory, we then need to average Eqs.\,\eqref{Average} and \eqref{Apure} over all possible values of $\phi$ [or, equivalently, consider a mixed state with $\vec{n}=(0,0,n \cos\theta$)], resulting in
\begin{align}
\!\!\frac{ \overline{\langle\hat{a}\rangle}_{\rho}}{g} \ce \!\int_{-\pi}^{\pi} \!\!d\phi \, \frac{ {\langle\hat{a}\rangle}_{\rho}}{g}&= 
\frac{r_1 + r_2}{2}  +  \frac{r_1-r_2}{2} n \cos \theta ,
\label{AvPhiA} 
\end{align}
\begin{align}
\frac{ \overline{ (\Delta \hat{a}^2)}_\rho}{g^2} \ce \!
\int_{-\pi}^{\pi} \!\!d\phi \,\frac{(\Delta \hat{a}^2)_\rho }{g^2} &= 
\bigg(\frac{r_1 - r_2}{2}\! \bigg)^{\!2} ( 1 - n^2 \cos^2 \theta)\notag \\
&\quad +  |r|^2\bigg(\!1-\frac{n^2}{2} \sin^2 \theta \!\bigg). 
\label{AvPhi}
\end{align}
Even in the realistic case of $0 < n <1$, a comparative analysis of the measured variances for different values of $\theta$ (in particular, $\theta=0$ and $\theta=\pi/2$) allows for the determination of $r_1-r_2$ and $|r|$, without needing any knowledge of the mean value of the acceleration. In Appendix~A, we further discuss the shortcomings of using Eq.\,\eqref{Average} alone to provide bounds on $|r|$, as well as the robustness of the results using Eq.\,\eqref{AAverage} with respect to small errors in the initial state preparation.

The main takeaway of this section is that the {\em variance} of the acceleration is the key parameter for studying genuine quantum 
violations of the WEP in the setting of free-fall experiments, particularly if there is no control over the angle $\phi$ in the Bloch sphere. 
It is natural to extend this discussion to the setting of torsional balance configurations which is our primary focus, as we start addressing in the next Section.

\section{An E\"{o}tv\"{o}s test of the WEP for \\ quantum matter}
\label{EotvosSection}

We now apply the formalism discussed in Sec.\,\ref{Gravitational and inertial masses as operators} to potential tests 
of the WEP for quantum states of matter using torsion balances. In the original experiment of E\"{o}tv\"{o}s \cite{Eotvos}, 
the forces acting on each test body were the gravitational force sourced by the Earth and the centrifugal force due to the Earth's rotation about its axis. The former is proportional to the gravitational mass of the test body, the latter to its inertial mass. 
The resulting torques are time independent, and the amplitude of the torque due to the gravitational force is much larger than 
the one due to the centrifugal force, even in the optimal case of an experiment located at the equator. 

Alternatively, one can aim at detecting the torques exerted on the test body due to the gravitational force sourced by the Sun, and the centrifugal acceleration due to the revolution of the Earth about the Sun. In this case, the expected signal due to an unbalance of the torques is dependent on time with a daily periodicity, and the gravitational and inertial torques are comparable. This is the configuration used by Roll, Krotkov, and Dicke \cite{Roll1964}, as well as Braginsky and Panov \cite{Braginsky1971}, to achieve the best-to-date bounds on the WEP in a laboratory setting. Here, we revisit these experiments as possible tests of the WEP in the presence of quantum coherence. This requires a detailed description of the torque at the classical level, and thereafter its promotion to an operator, in order to consistently incorporate the assumed operator nature of both the gravitational and inertial masses.

\subsection{Torsional balance setting and torque operator}
\label{sub:Torque}

\begin{figure}[t]
\centering
\includegraphics[width=8.6cm]{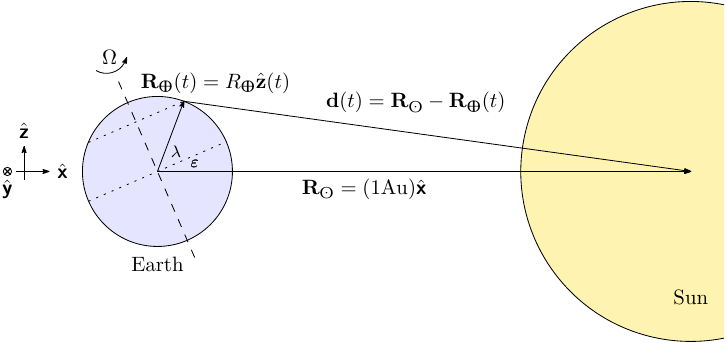}
\vspace*{-3mm}
\caption{Relative geometry between the Earth and Sun when the magnitude of ${\bf d}(t)$ is smallest and 
Earth is at the winter solstice in the northern hemisphere. 
An Earth-centered coordinate system is used in which the $z$ axis 
is orthogonal to the ecliptic plane and the Sun lies along the positive $x$ axis at the winter solstice in the northern hemisphere. 
The latitude of the torsion balance is $\lambda$ and $\varepsilon \approx 23.4^\circ$ is the axis tilt of the Earth relative to the ecliptic plane.}
\label{Fig:EarthSun}
\end{figure}

The torsion balance we consider is depicted in Fig.\,\ref{Fig:eotvos}, in the noninertial lab frame that rotates with the Earth, which is centered on the torsion balance and associated with the Cartesian unit vectors $\{\hat{\bf x},\hat{\bf y},\hat{\bf z}\}$. The positive $z$ axis is chosen to be opposite to the local gravitational field of the Earth and the $x$ axis points due south, so that the centrifugal force on each arm of the balance due to the Earth's rotation about its axis lies in the $xz$ plane. 
Labeling each arm of the balance with the index $j \in \{A,B\}$, the gravitational force from the Earth ${\bf F}_{\mathTerra,j}$ and the centrifugal force ${\bf F}_{\Omega,j}$ associated with the rotation of the Earth about its axis are expressed in the lab frame as
\begin{align*}
    {\bf F}_{\mathTerra,j} &= -M_{g,j} g_{\mathTerra} \hat{\bf z}, \\
    {\bf F}_{\Omega,j} &=  M_{i,j} \Omega^2 R_{\mathTerra} \cos \lambda \left( \sin \lambda \hat{\bf x}  +  \cos \lambda \hat{\bf z}\right),
\end{align*}
where $M_{i,j}$ and $M_{g,j}$ denote the inertial and gravitational mass of the $j$th balance arm, $g_{\mathTerra}$ is the local acceleration due to the Earth's gravitational field, $\Omega$ is the angular velocity of the Earth,  $R_{\mathTerra}$ is the radius of the Earth, and $\lambda$ is the latitude of the balance's location. The gravitational force of the Sun, ${\bf F}_{M_\mathSun,j}$, and the centrifugal force associated with the Earth's orbit about the Sun, ${\bf F}_{\bar{\Omega},j}$, on the balance are given by 
\begin{align*}
    {\bf F}_{\mathSun,j}(t)   &=   M_{g,j }\frac{G_N M_\mathSun}{d(t)^2} \hat{\bf d}(t), \\
    {\bf F}_{\bar{\Omega},j} &=  M_{i,j}  {R_\mathSun}\bar{\Omega}^2  {\hat{\bf R}_\mathSun},
\end{align*}
where $G_N$ is Newton's gravitational constant, $M_{\mathSun}$ is the mass of the Sun, $R_{\mathSun} = 1\,\text{Au}$ is the average distance between the Earth and Sun, and $\bar{\Omega}$ is the angular velocity of the Earth about the Sun. As shown in Fig.\,\ref{Fig:EarthSun}, the vector ${\bf d}(t) \ce{\bf R}_{\mathSun} - {\bf R}_{\mathTerra}(t)$ points from the torsion balance toward the Sun, $\hat{\bf R}_\mathSun$ points from the center of the Earth to the Sun and lies in the ecliptic plane, and ${\bf R}_{\mathTerra}(t)$ points from the center of the Earth to the location of the balance; for simplicity, we will take the orbit of the Earth about the Sun to be circular. The vectors $\hat{\bf R}_{\mathSun}$ and ${\bf d}(t)$ are depicted in Fig.\,\ref{Fig:EarthSun} relative to an Earth-centered frame $\{\hat{{\bf{\mathsf{x}}}},\hat{{\bf{\mathsf{y}}}},\hat{{\bf{\mathsf{z}}}}\}$ co-orbiting with the Earth about the Sun, in such a way that $\hat{\bf{\mathsf{x}}}$ always points toward the Sun. The net force acting on each arm of the torsion balance is thus given by
\begin{align}
{\bf F}_j &=  {\bf F}_{\mathTerra,j} + {\bf F}_{\Omega,j} + {\bf F}_{M_\mathSun,j} + {\bf F}_{\bar{\Omega},j} \nonumber \\
& \ce  {\boldsymbol{\beta}}(t) {M}_{i,j} +  {\boldsymbol{\gamma}}(t) M_{g,j }, \quad j\in\{A,B\},
\label{NetForce}
\end{align}
where the vectors ${\boldsymbol{\beta}}(t)$ and ${\boldsymbol{\gamma}}(t)$ are computed in Appendix~\ref{ComputationOfVectors}.

The torque about the intersection of the fiber and the balance arms is $\boldsymbol{\tau} = {\bf r}_A \times {\bf F}_A + {\bf r}_B \times {\bf F}_B$. Specifically, the torque that generates a twist of the balance is the component of $\boldsymbol{\tau}$ along the fiber
\begin{align}
\tau_{\parallel} &= \boldsymbol{\tau}\cdot \frac{{\bf T}}{\left| {\bf T} \right|} 
= -\left({\bf r}_A \times {\bf F}_A + {\bf r}_B \times {\bf F}_B \right) \cdot
\frac{{\bf F}_A + {\bf F}_B}{\left|{\bf F}_A + {\bf F}_B\right|}  
\nonumber \\
&= {\bf r}_{AB} \cdot \frac{{\bf F}_A \times  {\bf F}_B}{\left|{\bf F}_A + {\bf F}_B\right|}.
\label{signal}
\end{align}
Here, we have used the fact that, at equilibrium, the tension ${\bf T}$ in the fiber is balanced by the forces ${\bf F}_A$ and ${\bf F}_B$, whereby ${\bf T} +{\bf F}_A + {\bf F}_B = 0$; in the second line, we have defined the vector ${\bf r}_{AB} \ce {\bf r}_B - {\bf r}_A $ and made use of the cyclic property of the triple product, ${\bf a} \times {\bf b} \cdot {\bf c}= {\bf b} \times {\bf c} \cdot {\bf a}= {\bf c} \times {\bf a} \cdot {\bf b}$. It is precisely this self-cancellation of forces at equilibrium that provides an enormous advantage in using the torsional balance setup with respect to free-fall experiments, in spite of the far smaller accelerations present, especially in the Dicke-Bragingky configuration.

Substituting Eq.\,\eqref{NetForce} into Eq.\,\eqref{signal} yields
\begin{align}
\tau_{\parallel} &=  \frac{ {\bf r}_{AB} \cdot  {\boldsymbol{\gamma}}(t) \times {\boldsymbol{\beta}}(t) {M}_{i,A} {M}_{i,B}}
{\left| {\boldsymbol{\beta}}(t) \left( {M}_{i,A} + {M}_{i,B} \right) +  {\boldsymbol{\gamma}}(t) \left({M}_{g,A} + {M}_{g,B} \right) \right|}
 \nonumber \\
 & \quad  \times\left({M}_{g,A}{M}_{i,A}^{-1}  - {M}_{g,B}{M}_{i,B}^{-1}\right), 
\label{torque}
\end{align}
where ${\boldsymbol{\beta}}(t)$  and  ${\boldsymbol{\gamma}}(t)$ are related to the sources of gravitational and inertial forces as explicitly 
discussed in Appendix B, see in particular Eqs.\,\eqref{betaexpress} and \eqref{gammaexpress}. Upon taking the absolute value of the above expression and considering the definition of $\eta$ in Eq.\,\eqref{Eotvos}, it is seen that the  E\"{o}tv\"{o}s parameter $\eta$ is directly proportional to the torque about the fiber.

We now quantize the torque parallel to the fiber. In complete generality, this would be achieved by promoting all the relevant quantities present in Eq.\,\eqref{torque} to operators, including the distance ${\bf r}_{AB}$. However, we instead just replace the inertial and gravitational mass in Eq.\,\eqref{torque} by their postulated operator forms, $\hat{M}_{i,j} = m_{i,j} \hat{I} + \hat{H}_{i,j}/c^2$ and $\hat{M}_{g,j} = m_{g,j} \hat{I} + \hat{H}_{g,j}/c^2$, respectively. This is justified because of the {\it differential} character of the measurement we are going to propose, as discussed at the end of this Section, with the quantum fluctuations associated to the distance ${\bf r}_{AB}$, for a nearly rigid rod, being completely negligible.

Suppose now that we assume the validity of the {\em classical} WEP, $m_{g,j} = m_{i,j} = m_j$, for $j\in \{A,B\}$, which has been confirmed to a precision of $|m_g-m_i| \leq 10^{-15}$~\cite{Touboul2019a,Touboul2019b,Touboul2022}, far greater than what can be achieved in the torsion balance experiment we propose. Then, by substituting the mass operators into Eq.\,\eqref{torque} and retaining terms up to leading order in $1/c^2$, the torque operator takes the form (see Appendix~\ref{ComputationOfVectors})

\begin{equation}     \hat{\tau}_\parallel(t) =  \tau_0(t) \bigg(  \frac{\hat{H}_{g,A} - \hat{H}_{i,A}}{m_{A}c^2} - \frac{\hat{H}_{g,B} - \hat{H}_{i,B}}{m_{B}c^2}\bigg)+ \mathcal{O}\!\left(1/c^4\right),
     \label{torque2}
\end{equation}
where we have defined
\begin{equation}
\tau_0(t) \ce \mu \,\frac{ {\bf r}_{AB} \cdot  {\boldsymbol{\gamma}}(t) \times {\boldsymbol{\beta}}(t)}{ \left|{\boldsymbol{\beta}}(t) + {\boldsymbol{\gamma}}(t)\right| } , \quad \mu \ce \frac{m_A m_B}{ m_A+m_B} .
\label{torqueFactor}
\end{equation}
For simplicity, in what follows we will further assume that each arm of the balance consists of the same amount of material, so that $m_A = m_B = m$,  hence $\mu = m/2$.

\subsection{Mean and variance of the torque operator} 

To make our proposal more concrete, let us imagine that each arm of the balance consists of $N$ (approximately) noninteracting two-level quantum systems, so that the inertial and gravitational Hamiltonians $\hat{H}_{i,j}$ and $\hat{H}_{g,j}$ of each arm, acting on ${\cal H}_j\simeq ({\mathbb C}^2)^{\otimes N}$, may be taken to be the sum of $N$ two-level (identical) Hamiltonians. That is, with $\zeta \in \{i,g\}$,  $\hat{H}_{\zeta,A} = \big(\sum_{l=1}^N \hat{H}_{\zeta}^{(l)}\big) \otimes \hat{I}_B$, with $\hat{I}_B$ denoting the identity operator on ${\cal H}_B$ and $\hat{H}_{\zeta}^{(l)} = (\otimes_{k=1}^{l-1} \hat{I}) \otimes \hat{H}_{\zeta} (\otimes_{k=l+1}^N \hat{I})$ acting nontrivially only on the $l$th subsystem, with $\hat{I}$ denoting as before the identity on ${\mathbb C}^2$. Likewise, $\hat{H}_{\zeta,B} = \hat{I}_A\otimes \big(\sum_{l=1}^N \hat{H}_{\zeta}^{(l)}\big)$. Thus, for arm $j=A$, the difference 
\begin{equation}
    \frac{\hat{H}_{g, A} - \hat{H}_{i,A} }{mc^2} =
 \sum_{ l=1}^N   \begin{pmatrix}
        r_1 -1 & r \\
        r^* & r_2 - 1
    \end{pmatrix}^{\!\!(l)} \otimes \hat{I}_B,  
\label{HamiltonianAndr}
\end{equation}
in terms of the matrix elements $r_1$, $r_2$, and $r = |r| e^{i\varphi_r}$ introduced in Eqs.\,\eqref{aFreeFall} and \eqref{aApprox}, with a similar expression holding for $j=B$. Further, suppose that the quantum state of each arm of the balance, $\hat{\rho}_A$ and $\hat{\rho}_B$, is an $N$-fold tensor product of the state in Eq.\,\eqref{TwoLevelState}, 
\begin{equation}
    \hat{\rho}_j = 
    \hat{\rho}(n_j,\theta_j,\phi_j)^{\otimes N} , 
    \label{NfoldTP}
\end{equation}
with the joint state of the two arms assumed to be factorized, $\hat{\rho}= \hat{\rho}_A \otimes \hat{\rho}_B$ on ${\cal H}_A\otimes {\cal H}_B.$
Then, the expectation value of $\hat{\tau}_{\parallel}$ in Eq.\,\eqref{torque2} is found to be 
\begin{align}
\frac{ \langle{\hat{\tau}}_{\parallel} \rangle_{\rho}  }{\tau_0(t)} 
&\ce N\big[ F(r_1,r_2,|r|,\varphi_r; n_A,\theta_A,\phi_A) \nonumber \\
&\qquad \quad  - F(r_1,r_2,|r|,\varphi_r; n_B,\theta_B,\phi_B)\big]\nonumber \\
&= N \bigg( \frac{r_1-r_2}{2} \big[n_A \cos \theta_A - n_B \cos \theta_B \big] \nonumber \\ 
&\qquad \quad  +|r| \cos \varphi_r   \big[ n_A \cos (\phi_A-\varphi_r) \sin \theta_A \nonumber \\
&\qquad \quad  
 -n_B \cos (\phi_B-\varphi_r) \sin \theta_B \big] \bigg), \nonumber 
\end{align}
in analogy to Eq.\,\eqref{Average}. Similarly,  its quantum variance reads  
\begin{eqnarray*}
({ \Delta {\hat{\tau} }_\parallel^2})_{\rho} & = & {\braket{\hat{\tau}_\parallel^2}_\rho} -{\braket{\hat{\tau}_\parallel}}_\rho^2  = 
( \Delta \hat{\tau}^2_A )_\rho + (\Delta \hat{\tau}^2_B)_\rho ,\\
\frac{ ( \Delta \hat{\tau}^2_j )_\rho }{\tau_0(t)^2} & \ce & N \, G(r_1,r_2, |r|, \varphi_r ; n_j,\theta_j, \phi_j) , 
\end{eqnarray*}
where we have used the additivity of the variance for a product state and the function $G$ has the expression given in Eq.\,\eqref{AAverage}. Thus, the mean and variance of the torque operator depend on the initial preparation of the quantum state in each arm via $n_j$, $\theta_j$, and $\phi_j$. 
Notice that, unlike the free fall case discussed in Sec.\,\ref{Gravitational and inertial masses as operators}, a maximally mixed state for both test masses ($n_A=n_B=0$) does not produce any signal even in the presence of nonzero $r_1$ and $r_2$, due to the differential nature of the torque observable. The choice $\theta_B =\pi-\theta_A$ maximizes both the mean torque and its variance and, under this assumption, considering the two initializations $\theta_A=0$ and  $\theta_A=\pi/2$ allows for separate bounds on $r_1-r_2$, or $r$ in principle. 

Similar to the free-fall case, averaging over the random phases $\phi_A$ and $\phi_B$ results in a simplified expression for the average torque,
\begin{equation*}
\overline{ 
\frac{ \langle {\hat{\tau}_\parallel \rangle}_\rho} {\tau_0(t)} } =\! \int_{-\pi}^\pi 
\!\! \!d\phi \, \frac{\braket{\hat{\tau}_{\parallel}}}{\tau_0(t)} = N\frac{r_1-r_2}{2} \left(n_A  \cos \theta_A - n_B \cos\theta_B\right) ,
\end{equation*}
which is the analog of Eq.\,\eqref{AvPhiA}. Likewise, the average variance of the torque is given by 
$$ \overline{\frac{ ({ \Delta {\hat{\tau} }_\parallel^2})_{\rho} }{\tau_0(t)^2} }= \! \int_{-\pi}^\pi 
\!\! \!d\phi \, 
\frac{ ( \Delta \hat{\tau}^2_A )_\rho}{\tau_0(t)^2}   +  \int_{-\pi}^\pi 
\!\! \!d\phi \,  \frac{(\Delta \hat{\tau}^2_B)_\rho}{\tau_0(t)^2}, $$
with each of the above averages formally resulting in the same expression as in Eq.\,\eqref{AvPhi}. As for free fall, the optimal case would correspond to a pure state, with $n_A=n_B=1$, and, as noted above, $\theta_A+\theta_B=\pi$.

In an actual experimental setup, the total variance will also include a purely {\em classical} contribution; in a properly designed experiment the most significant classical contribution (and easiest to model) is due to the thermal motion of the apparatus. Ensuring that the thermal contribution is reduced below the quantum variance is important for determining the ultimate bounds on $r_1, r_2,$ and $r$ we discussed above. Even assuming that this is achievable, a main challenge arises from the need to prepare the $N$ two-level systems that comprise each of the two test masses in highly coherent, ideally pure quantum superposition states, and preserve them for the duration of the experiment. Given the need to operate at ambient temperature, high-density NV-center ensembles appear especially promising, thanks to the unique combination of enhanced signal-to-noise ratio (SNR), high degree of control, and spin-state readout they can offer \cite{NVReview}. Several noise sources, however, ranging from quasistatic field or strain inhomogeneity to spin-bath couplings, contribute to the ensemble dephasing time, $T_2^*$, and the single-spin coherence time, $T_2$; while, in a nitrogen-rich sample, the latter timescale can be up to two orders of magnitude longer than the dephasing time $T_2^*$, both are severely degraded with increasing defect concentration, with inverse linear scalings having been reported \cite{NVReview,Baugh}. 

In principle, both $T_2^*$ and $T_2$ may be substantially prolonged by the use of suitable multipulse dynamical decoupling (DD) methods \cite{ViolaDD,SlavaDD,Khodjasteh2013}, with recent studies pointing to the importance of a fully quantum treatment \cite{Galli}. 
Timescales of milliseconds might be eventually realizable (remarkably, record-long coherence times on the scale of minutes have been recently reported through the use of DD in a single silicon-carbide defect center \cite{Awschalom}). However, such timescales are still realistically too far from the typically long timescales of torsional balances. The theory we have described, and most notably the self-balancing of the torsional balance, occurs on a long timescale, on the order of minutes to hours. 

To overcome this challenge, two possible paths are worth mentioning. 
First, the use of specialized quantum-control techniques may allow for high-fidelity preservation of a {\em designated} quantum state for timescales much longer (arbitrarily long in principle) than achievable with general-purpose DD schemes \cite{Khodjasteh2011}. If, for instance, both arms of the balance are prepared in an optimal pure state of the form $|\psi_+\rangle_j = \tfrac{1}{\sqrt{2}}( |1\rangle_j + |2\rangle_j)$, $j=A,B$, a ``ZZ'' pulse sequence that periodically exchanges the phase between the two energy eigenstates (hence maps $|\psi_+\rangle \mapsto |\psi_-\rangle$) may be used to effectively freeze the quantum state into an approximate ``pointer state'' for the dynamics, with long-time fidelities significantly higher than achievable with a standard two-axis (e.g., ``XYXY'') DD scheme. While continual improvements in coherent-control capabilities and coherence times, together with progress in the miniaturization of torsional balances (see for example Ref.\,\cite{Lami2024}), could thus bring our proposed test closer to feasibility, a second\,---\,ultimately more compelling\,---\,path is to turn the attention to torsional balances in which the source of the gravitational field is {\em time dependent}. This may be seen as a compromise between the free-fall setup and the usual torsional balance, in that the self-adjusted equilibrium of the latter is no longer ensured, yet, in a way, the two test bodies ``keep falling'' on the gravitational source, as we proceed to describe in the following Section.

\section{A dynamical Cavendish test of the WEP for quantum matter: A Cavend\"{o}s experiment}
\label{Dynsetup}

As an alternative to the E\"{o}tv\"{o}s-type experiment described in the previous Section, the analysis of which is based on the assumption of a static gravitational field, let us consider the response of a torsion balance to a {\em dynamical} gravitational field. Time-dependent sources of gravitational fields have been demonstrated since the inception of gravitational-wave research, using vibrating masses via piezoelectric driving, and creating in this way a quadrupolar dynamical gravitational field \cite{Sinsky1967,Sinsky1968}. Thereafter, the source of the dynamical gravitational field was chosen to be a rotating mass, and more detailed experimental studies of the near-field region have been performed in the context of searching for possible deviations from the inverse-square law \cite{Hirakawa1980,Astone1991,Astone1998}. More recently, rotating masses have also been designed to specifically calibrate gravitational wave interferometers \cite{Ross2021,Ross2023}.

\subsection{Dynamical torsion balance setup}

\begin{figure}[t]
 \centering
 \includegraphics[width=8cm]{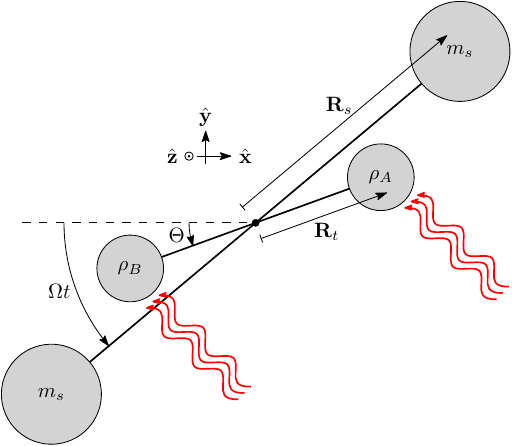}
 \caption{Schematics for the proposed dynamical Cavendish experiment. As in the static case, the internal states of the balance's arms 
 are described by density operators $\hat{\rho}_A$ and $\hat{\rho}_B$; in addition, two identical sources with mass $m_s$ generate 
 time-dependent gravitational fields ${\bf g}_A(t)$ and ${\bf g}_B(t)$ [Eqs.\,\eqref{gA}--\eqref{gB}] by rotating with a constant angular frequency $\Omega$. Here, $\Theta$ represents the average angular displacement of the balance referred to the $x$ axis, and by construction we have ${\bf R}_t = (R_t \cos \Theta, R_t \sin \Theta, 0)$, $ {\bf R}_s= (R_s \cos \Omega t, R_s \sin \Omega t, 0)$.  The red wavy lines represent the laser pulses required to prolong the coherence time of the two test masses, and possibly dynamically freeze them in their target initial state. }
\label{fig:Cavendish}
\end{figure}

Based on the achievements described above, we consider a setup in which the dynamical gravitational field is sourced by two spheres of mass $m_s$ that rotate at an angular speed $\Omega$ in the same plane of the arms of the torsional balance, as depicted in Fig.\,\ref{fig:Cavendish}. The torque about the balance due to the rotating masses, when the classical WEP holds, $m_{i,j} = m_{g,j} = m$ for $j\in\{A,B\}$, is now characterized by the time-dependent operator 
\begin{equation}
    \hat{\tau}(t) = \left[ {\bf R}_s \times {\bf g}_A(t) \hat{M}_{g,A} - {\bf R}_s \times {\bf g}_B(t) \hat{M}_{g,B} \right] \cdot\hat{{\bf z}},
    \label{dynamicalTorque}
\end{equation}
where ${\bf g}_j(t)$ is the instantaneous gravitational field at each arm of the balance and ${\bf R}_s$ points from the midpoint of the balance to a test mass as depicted in Fig.\,\ref{fig:Cavendish}.
The idea is that ${\bf g}_j(t)$  will vary on a characteristic timescale much shorter than considered in Sec.~\ref{EotvosSection}, $\hat{\tau}_\parallel \propto \tau_0(t)$,  
for which the variation is due to the motion of the Earth about its axis and around the Sun; thus, any contribution to the torque due to quantum coherence between internal energy levels that varies on this shorter timescale can serve as a test of the WEP for quantum matter. 

Unlike the E\"{o}tv\"{o}s-type experiment discussed in Sec.\,\ref{EotvosSection}, the torsion balance will not be in equilibrium, given that ${\bf g}_j(t)$ is assumed to vary on a timescale much shorter than the response time of the balance. Thus, the torque will no longer be proportional to the angular displacement of the balance. Instead, the relevant observable will be the {\em angular acceleration} of the torsion balance due to $\hat{\tau}$, 
\begin{equation}
\hat{\alpha}(t) \ce {\hat{\mathcal{I}}}^{-1} {\hat{\tau}(t)}, \quad \hat{\mathcal{I}} =|{\bf R}_t |^2 (\hat{M}_{i,A} +   \hat{M}_{i,B} ),
\label{eq:angacc}
\end{equation}
where $\hat{\mathcal{I}}$ is the moment of inertia operator  of the balance. 
Specializing to the geometry in Fig.\,\ref{fig:Cavendish}, note that the gravitational field at $A$ and $B$ are given by 
\begin{eqnarray}
\frac{{\bf g}_A(t)}{G_N m_s}  
&=& \frac{ +{\bf R}_t - {\bf R}_s  } {{|{+\bf R}_t - {\bf R}_s|}^3} 
+ \frac{ +{\bf R}_t + {\bf R}_s } {|+ {\bf R}_t + {\bf R}_s|^3} , 
\label{gA} \\
\frac{{\bf g}_B(t)}{ G_N m_s}  
&=& \frac{-{\bf R}_t - {\bf R}_s }{|{-{\bf R}_t - {\bf R}_s|}^3} +
\frac{ -{\bf R}_t + {\bf R}_s}{|-{\bf R}_t + {\bf R}_s|^3}
 \label{gB}.
\end{eqnarray}
Let $R_s\ce|{\bf R}_s|$, $R_t\ce|{\bf R}_t|$ and let us also introduce the shorthand notation
\begin{align}
R_{\pm}(t) &\ce|{\bf R}_t\pm {\bf R}_s| \nonumber \\
&= \sqrt{R_t^2 + R_s^2 \pm 2 R_t R_s \cos \left(\Omega t- \Theta \right)}.
\label{Carnot}
\end{align}
By assuming, as before, that each test mass consists of $N$ two-level systems, the angular acceleration due to the dynamical 
gravitational field to the leading order in $1/c^2$ may then be computed as follows (see Appendix \ref{CavendishAppendix} for detail): 
\begin{eqnarray}
    \hat{\alpha}(t) &\!\!= \!\!& \frac{G_N {m}_s}{2}\frac{R_s}{R_t}  \left(\frac{1}{{R}_{+}(t)^3} -\frac{1}{{R}_{-}(t)^3} \right)  
    \sin \left(\Omega t- \Theta \right)  \nonumber \\
& \!\!\times \!\!& \sum_{l=1}^N 
\left[ \begin{pmatrix}
r_1  & r \\
r^* &  r_2
\end{pmatrix}^{\!\!(l)} \!\!\!\otimes \hat{I}_B + \hat{I}_A \otimes \begin{pmatrix}
r_1  & r \\
r^* &  r_2
\end{pmatrix}^{\!\!(n)} \right].
\label{angularAcceleration}
\end{eqnarray}
Accordingly, the angular acceleration $\hat{\alpha}$ is sensitive to the parameters $r_1$, $r_2$, and $r$.

Suppose that the arms of the balance are both prepared in the same product state $\hat{\rho}_A = \hat{\rho}_B = \hat{\rho}$, given 
in Eq.\,\eqref{NfoldTP}. The mean and the quantum variance of the angular acceleration operator are then obtained as follows:
\begin{align}
\langle{\hat{\alpha}(t)}\rangle_\rho&\!= \!N G_N m_s\frac{R_s}{R_t}  
\left(\frac{1}{{R}_{+}(t)^3} -\frac{1}{{R}_{-}(t)^3} \right)  \nonumber \\
&\quad \times  \sin \left(\Omega t- \Theta\right) F(r_1,r_2,|r|,\varphi_r;n,\theta,\phi), 
\label{Acceleration} \\
(\Delta \hat{\alpha}^2(t))_\rho &\!= \!N \left[G_N m_s \frac{R_s}{R_t} 
\left(\frac{1}{{R}_{+}(t)^3} -\frac{1}{{R}_{-}(t)^3} \right)\right]^2   \nonumber \\
&\quad \times \sin^2 \left(\Omega t-\Theta \right) G(r_1,r_2,|r|, \varphi_r;n,\theta,\phi).
\label{Variance}
\end{align}
Equations \eqref{Acceleration} and \eqref{Variance} contain the main message of this section, and we now discuss 
various cases, keeping in mind that  $r_1$, $r_2$, and $r$ parametrize violations of WEP and, as such, cannot be controlled, while others, $n, \theta, \phi$, characterize the state preparation of the setup\,---\,with all of them affecting the values of the form factors $F$ and $G$ as in Eqs.~\eqref{Average} and \eqref{AAverage}. 

In the case where the WEP holds, we have $r_1=r_2=1$ and $r=0$, leading to $F=1$ and $G=0$ {\em regardless} of the state 
preparation. The acceleration will then reduce to the one expected classically, 
\begin{align}
\langle{\hat{\alpha}(t)}\rangle_{\rho}&= N G_N m_s \frac{R_s}{R_t}  
\bigg(\frac{1}{{R}_{+}(t)^3}-\frac{1}{{R}_{-}(t)^3} \bigg) \sin \left(\Omega t- \Theta\right)  \nonumber \\
& 
=: \alpha_{\mathrm{cl}}(t) ,
\label{PurelyClassical}
\end{align}
with a zero variance quantum $(\Delta \hat{\alpha}^2(t))_\rho$ related to the system's state. 

Even in the absence of a variance depending on $n, \theta, \phi$, however, we expect uncertainty related to both the influence of the environment on the torsional balance, in the form of thermal Brownian motion as mentioned before, as well as due to imperfect 
knowledge of all the parameters appearing in Eq.\,\eqref{PurelyClassical}, most notably the precision in the measurement of $G_N$, currently known at the level of 47 parts per million \cite{Xue}. This variance, being of a purely classical origin, will be denoted as $\Delta \alpha^2_{\mathrm{cl}}$ in what follows. 

As a first case where a quantum contribution might manifest, we consider a situation where $r \neq 0,  n=0$, and maintain $r_1=r_2$. The mean angular acceleration in Eq.\,\eqref{Variance} is unaffected, but now there is a new contribution to the variance, which can be written as 
\begin{equation}
(\Delta \hat{\alpha}^2(t))_\rho = \frac{1}{N} 
 \alpha_{\mathrm{cl}}(t)^2  |r|^2 + \Delta \alpha^2_{\mathrm{cl}}.
\label{Variance1}
\end{equation}
Therefore, the variance will be sensitive to the presence of off-diagonal matrix elements in the mass operators provided that the first term in the right-hand side of Eq.\,\eqref{Variance1} dominates with respect to the second, classical term. This occurs, for a fixed value of $|r|$, with a quadratic dependence on the acceleration signal. Notice that, for a given value of this signal (also depending on $N$), the inverse dependence on the number of ``active gravitational sources'' of quantum matter,  at least in the case in which the variance is linear in $N$.

A richer scenario occurs when $n \neq 0$. In this case, both the average angular acceleration and its variance are modified with respect to their classical values. Again, letting $r_1=r_2$ to isolate the off-diagonal effect, the maximum violation of the WEP occurs for pure states ($n=1$), in which case we have
\begin{align}
\langle{\hat{\alpha}(t)}\rangle_\rho &\!=\!
\alpha_{\mathrm{cl}}(t) \big[1+|r| \cos(\varphi_r+\phi) \sin \theta\big], 
\label{SignalPure} \\
(\Delta \hat{\alpha}^2(t))_\rho&\!=\! \frac{1}{N}  \alpha_{\mathrm{cl}}(t)^2 |r|^2 \big[1\!-\!\cos^2(\varphi_r+\phi)\sin^2 \theta\big]+ \Delta \alpha^2_{\mathrm{cl}}.
\label{VariancePure}
\end{align}
This is also the most advantageous case, because when the coherence between the internal energy eigenstates is maximized, 
$\theta=\pi/2$, the values of $|r|$ and $\varphi_r$ may be separately inferred from access to both the average acceleration and its variance, as long as control over $\phi$ is available. In all the situations with  $0 < n < 1$, we expect intermediate behavior between the two limiting cases discussed above, as pictorially shown in Fig.\,4 and also further discussed in the next Section. Similarly, if only partial control over the initial quantum state is available and averaging over $\phi$ is needed, weaker bounds may be expected, analogous to the static case. 

\subsection{Feasibility considerations and expected bounds}

A realistic description of the dynamics should take into account that the coherence of the state, apart from 
the case of $n=0$, is time dependent, and possibly rapidly decaying. Since, in this dynamical setting, the relevant timescales 
are much shorter than in the previously discussed DC torsional balance (e.g., on the order of tens of ms at 100 Hz), the use of DD or pointer-state engineering schemes becomes possible with existing or near-term capabilities.  However, another element one should explicitly account for is the time dependence of $R_+$ and $R_-$, which is dictated by the Newton equations for the angular motion of the test masses. 

\begin{figure*}
\includegraphics[width=18cm]{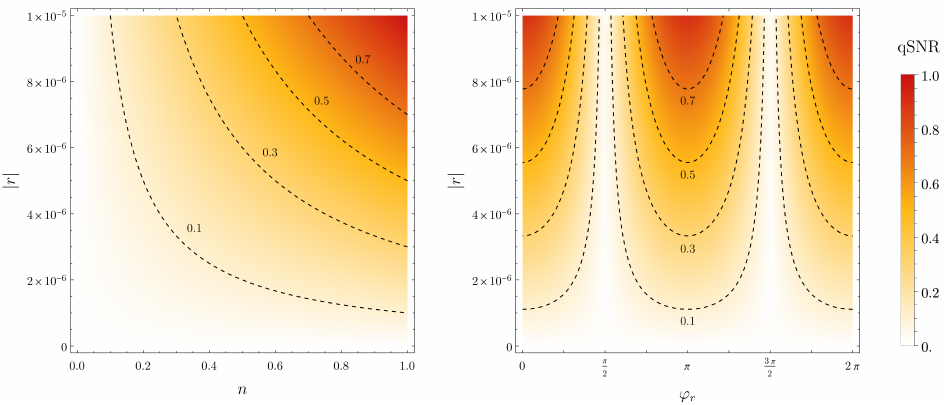}
\vspace*{-2mm}
\caption{Bounds on the off-diagonal matrix element $r$ in the regime where thermal fluctuations are small compared to the quantum variance of the acceleration operator, evaluated for $N = 10^5$ and $\sqrt{\Delta \alpha_{\mathrm{cl}}^{2}} /\alpha_\mathrm{cl}(t) = 10^{-5}$, close to the constraint set by the current relative precision in the measurement of $G_N$, with the dashed lines representing contours of constant $\mathrm{qSNR}$ as defined in Eq.\,\eqref{SNRQ}.   (Left) Dependence of $|r|$ upon the magnitude of the Bloch vector $n$ for different values of the $\mathrm{qSNR}$, in the optimized case of $\theta = \pi/2$ and $\varphi_r + \phi  = 0$.  Notice that the $\mathrm{qSNR}$ is a monotonically increasing function of both $n$ and $|r|$. For a given $n$, if a signal is not detected for a given $\mathrm{qSNR}$, all values larger than the corresponding $|r|$ in the vertical axis are ruled out.  (Right)  Contour plot of the $\mathrm{qSNR}$ in the $\varphi_r-|r|$ plane for the choice of $n=0.9$, $\theta = \pi/2$, and $\phi=0$.}
\label{FigSNR}
\end{figure*}

While a concrete experimental setup will deserve a full numerical analysis of the dynamics, we discuss here some general features. 
Regarding the time dependence of $R_+$ and $R_-$, useful insight may be gained in the limit where both the test masses are much closer to the fiber of the torsion balance than the source masses, that is, $R_t/R_s\ll 1$.  Equation \eqref{Carnot} can then be Taylor expanded to the leading order in this ratio, and Eqs.\,\eqref{Acceleration} and \eqref{Variance} can 
be, respectively, approximated as 
\begin{align}
    \langle{\hat{\alpha}(t)}\rangle_\rho &\approx -\frac{3NG_Nm_s}{R_s^3} \sin \left( 2 \Omega t - 2\Theta \right) 
    \nonumber \\
    &\quad 
    \times F(r_1,r_2,|r|,\varphi_r ; n,\theta,\phi), 
    \label{AccelerationApprox} 
\end{align}
and 
\begin{align}
(\Delta \hat{\alpha}^2(t))_\rho & \approx \frac{9 N G_N^2 m_s^2}{R_s^6} \sin^2 \left(2\Omega t-2\Theta \right) \nonumber \\
&\quad \times G(r_1,r_2,|r|,\varphi_r ; n,\theta,\phi).
\label{VarianceApprox}
\end{align}
The variance of the angular acceleration in Eq.\,\eqref{Variance}, which constitutes the signal of interest present even in the case of $n=0$, $r_1=r_2$, should be compared to the classical source of variance $\Delta \alpha_{\mathrm{cl}}$ as in Eq.\,\eqref{Variance1} and Eq.\,\eqref{VariancePure}. A torque sensitivity spectral density of $2 \times 10^{-17}$ N m/$\sqrt{\mathrm {Hz}}$  at 100 Hz, with a torsional oscillator of mass 10 mg, 
and dimensions 15 mm $\times$ 1.5 mm $\times$ 0.2 mm (moment of inertia around the fiber equal to $2 \times 10^{-10}$ Kg m${}^2$), has been reported \cite{Komori2020}. If this experiment could be performed maintaining the same noise level, with an integration time of 9 hours, we would achieve a minimum detectable $G$ in Eq.\,\eqref{Variance} of 
$G_{ \mathrm{min}}= 1.6 \times 10^{-6}$ by using gravitational sources with masses of 1 kg and $R_s=0.2$ m.  In optimal conditions, this would translate into a bound on $|r| \leq  4 \times 10^{-3}$. 

If we further assume that all technical noise has been suppressed\,---\,in particular, by maintaining acoustic and electrostatic insulation 
between the source masses and the test masses and by employing ideal detectors for the angle deflection signal\,---\,a remaining dominant source of noise would be the thermal noise on the test masses, as also mentioned before. Assuming that the noise is zero mean, this may be characterized by the torque Brownian noise spectral density, 
\begin{equation}
S_{\tau}^{\mathrm{th}} (\omega) = 4 \gamma I (\omega) k_B T,
\label{ThermalNoise}
\end{equation}
where $\gamma$ is the damping rate of the torsional mode, $I$ its moment of inertia, and $T$ the operating temperature. 
By assuming to work at sufficiently low frequencies $\omega$ with respect to the torsional mode frequency $\omega_m$, 
the damping rate is $\gamma=\omega_m^2/(\omega Q)$, with $Q$ being the mechanical quality factor of the torsional mode. 

From this, one may estimate a thermal torque spectral density of $8 \times 10^{-19}$ N m/$\sqrt{\mathrm {Hz}}$  at 100 Hz,  i.e., 25 times smaller than the sensitivity reported in \cite{Komori2020}. This translates into a gain factor of 25 in the bound on $|r|$, which would then become $|r| \leq 1.6 \times 10^{-4}$. 

Further sensitivity improvements are 
expected if the torsional balance is operated at cryogenic temperature, as in \cite{Fleischer2022}, due to both a reduction by two orders of magnitude in the temperature, and an unknown but likely significant increase in the mechanical quality 
factor of the torsional mode \cite{Bantel2000,Newman2014}, and using laser cooling techniques \cite{Agafonova2024a,Agafonova2024b,Shin}.
Notably, at cryogenic temperatures significantly longer coherence times may also be expected \cite{NVReview}, making it possible to use shorter integration times. 
Under such conditions, all the classical sources of noise would be smaller than the quantum uncertainty $(\Delta \hat{\alpha}^2(t))_\rho$ we wish to measure.

A useful indicator of the torque attributable to violations of the WEP arising from a nonzero $r$ may be obtained by introducing the following quantity:
\begin{align} 
\mathrm{qSNR} 
&\ce\frac{|\langle{\hat{\alpha}(t)}\rangle_\rho - {\alpha}_{\mathrm{cl}}(t) |}{\sqrt{ (\Delta \hat{\alpha}^2(t))_\rho + \Delta \alpha_\mathrm{cl}^2 }} 
\label{SNRQ} 
\\&=  \frac{ n | r |  \, |\cos \left(\varphi_r + \phi\right) \sin \theta | }{\sqrt{ \frac{|r|^2}{N} \Big(  1 -n^2  \cos^2 \left(\varphi_r + \phi\right) \sin^2 \theta  \Big) + \frac{ \Delta {\alpha}^2_{\mathrm{cl}}  }{{\alpha}_{\mathrm{cl}}(t)^2  }  }}.\notag
\end{align}
By construction, ${\mathrm{qSNR}}$ vanishes when the state is completely mixed, $n=0$; it is also zero when the state of the balance arms is in an energy eigenstate, $\theta \in \{0, \pi\}$, or the relative phase is tuned such that $\phi + \varphi_r\in \{\pi/2, 3\pi/2\}$.  Instead, the $\mathrm{qSNR}$ reaches its maximum when the states of the test masses are pure, $n=1$, equally weighted superposition of internal energy eigenstates, $\theta = \pi/2$, and the relative phase is tuned such that $\phi + \varphi_r \in \{0, \pi\}$. Therefore, if control over $\phi$ is available, one can change the sensitivity to different values of $\varphi_r$ in principle. This suggests that in a potential experiment, $\phi$ should be swept over to probe the entire parameter space with maximum sensitivity. Also note the $\sqrt{N}$ enhancement that stems from the use of a large number of ``active'' masses, provided that the quantum variance dominates.

These various dependencies are illustrated in Fig.\,4, where the behavior of the ${\mathrm{qSNR}}$ upon the degree of purity $n$ is shown for a variety of values of $\varphi_r+\phi$, $\theta$, and $|r|$. Here, we have assumed complete control over all the parameters of the Bloch sphere, so that arbitrary initial states can be prepared and maintained for the required duration, with no decoherence. Then any measurement of the $\mathrm{qSNR}$ in these conditions provides a combined bound on $|r|$ and $\varphi_r$. 

\section{Conclusions}

\begin{figure*}[t]
\includegraphics[width=8cm]{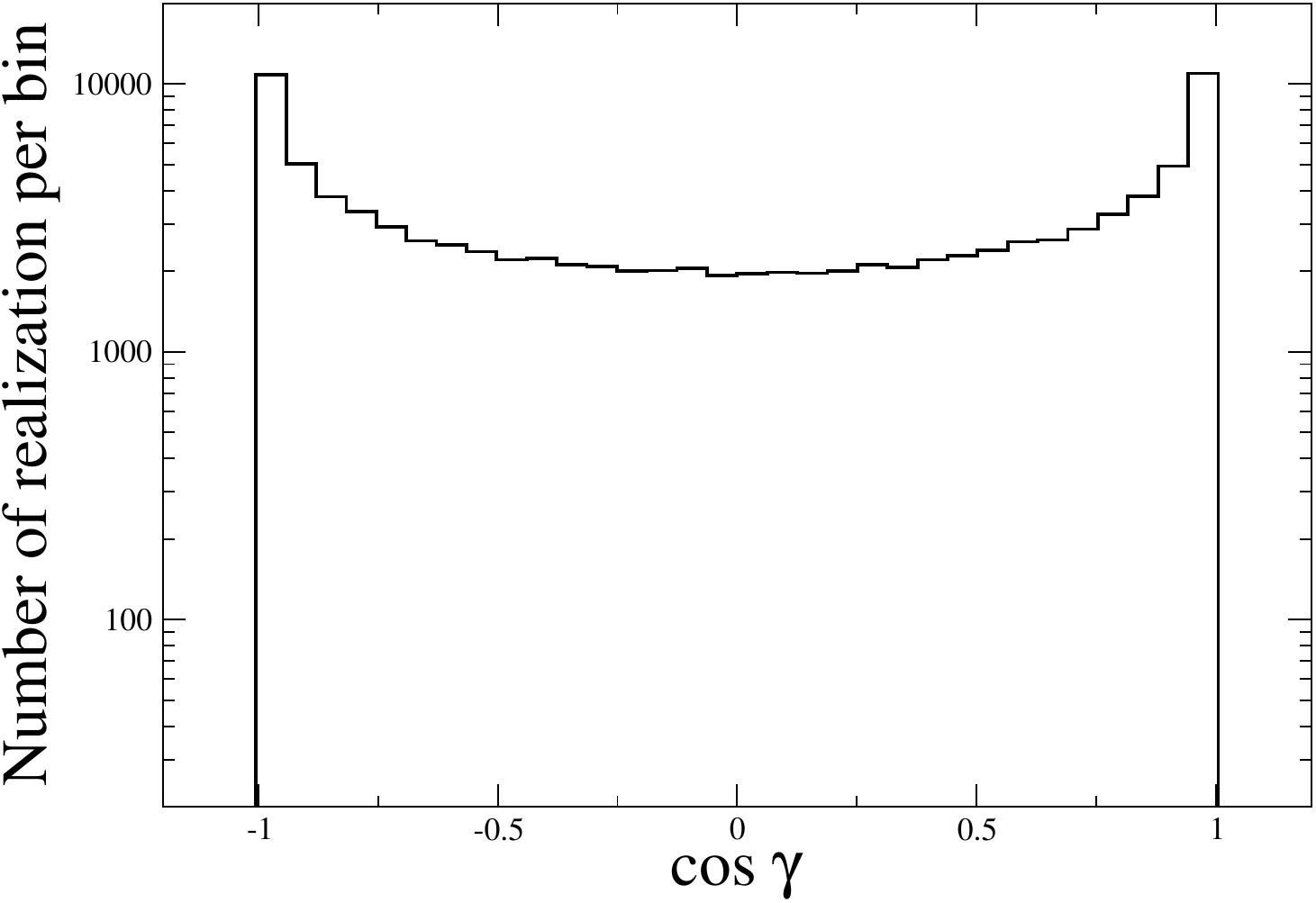}
\includegraphics[width=8cm]{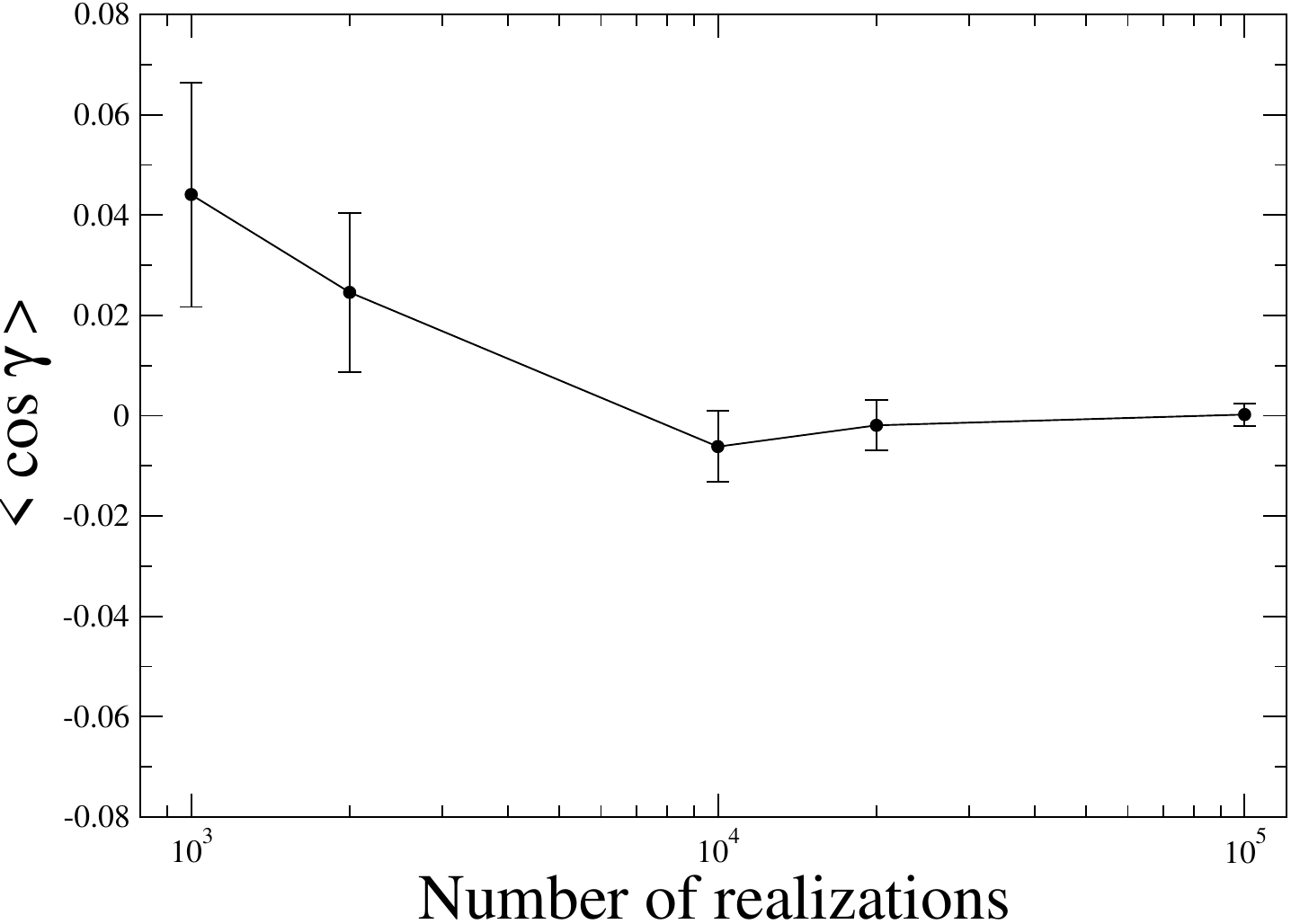}
\caption{(Left) Distribution of values of $\cos \gamma$ with $10^5$ realization of uniformly distributed real numbers for $\gamma\in  [-\pi, +\pi]$. (Right) Average of $\cos \gamma$ and its standard deviation versus the number of realizations, showing that its average value is 
compatible with zero within two standard deviations already for $10^3$ realization, and quickly converges to zero, reaching 
$\langle \cos \gamma \rangle =(2.3 \pm 22.4) \times 10^{-4}$ for $10^5$ realizations.}
\label{FigAppendix}
\end{figure*}

Building on a recent formulation of a quantum counterpart to the WEP in terms of suitable mass operators, we have discussed 
possibilities for precision tests based upon the use of torsional balances. We have shown that the parameters of the mass matrix are encoded in the measurable values of the average acceleration and the related variance, calculated for a generic quantum state, not necessarily with maximal purity. We have first discussed the simplest case of constraints based on free-fall experiments, which are plagued by the need to release the test masses at the same time with exquisite precision. Next, we have discussed the case of a torsional balance as in the E\"{o}tv\"{o}s and Dicke-Braginsky configurations. 
We have then discussed the possibility of a dynamical experiment intermediate between Cavendish and E\"{o}tv\"{o}s  configurations. This proposal uses a torsional balance under the action of a dynamical source of gravitational field, allowing for ac detection including classical and quantum \cite{Zhao} lock-in techniques, and the possibility to further modulate the effect we are aiming at by reviving at regular times the quantum coherence of the test mass.

Recent technological progress, such as the generation of dynamical gravitational fields, optomechanical torsion pendula, and the preparation and control of quantum states, could make such a test possible in dedicated experiments integrating all these techniques, also exploiting achievements in the measurement of the gravitational force at the microscale \cite{Westphal2021,Brack2022}. The estimated bounds on possible violations of the WEP do not  yet appear competitive with those based on atomic interferometry, the latter technique already capitalizing on three decades of dedicated research and development. However, also exploiting configurations and schemes currently under development for parallel goals such as searching for hints of quantum gravity at the microscale \cite{Smetana,Yan}, this could close the gap and complement the existing bounds based on atom interferometry, adding reliability of the outcome due to the presence of completely different systematic errors. The proposal could also greatly benefit from the preparation of the atoms in the two arms of the torsional balance in more general (possibly pairwise-entangled) quantum states, exploiting the growing body of work at the interface between quantum metrology and gravitational physics \cite{Aspelmeyer,Montenegro,Marletto,Bose}.

\subsection*{Acknowledgements}

R.O. gratefully acknowledges useful discussions with Giuseppe C. La Rocca. R.O. and L.V. are also grateful to the late Precooh for her invaluable support. 

\section*{Data Availability}

The data that support the findings of this article are not
publicly available. The data are available from the authors
upon reasonable request.

\appendix

\section{Robustness of the bounds on the WEP}

In this Appendix, we comment on two relevant aspects of the existing bounds on violations 
of the WEP at the quantum level, with specific reference to \cite{Rosi2017}.

A first potential issue is the robustness of the result against small errors in the preparation of the superposition state. 
In particular,  the authors discuss the acceleration in free fall for the atoms in one of the two 
energy eigenstates $|1\rangle$ or $|2\rangle$ [Eqs.\,\eqref{r1} and \eqref{r2} in their contribution] and in an equal-weight linear superposition [Eq.\,\eqref{rdiag}], that is, $\theta=\pi/2$. To explore the sensitivity to small deviations from this angle, we evaluate $\langle \hat{a} \rangle_{|s\rangle}$ and $\langle \hat{a}^2\rangle_{|s \rangle}$ 
by considering a superposition state of the form 
\begin{equation}
|s \rangle =  \cos ( \theta /2) \ |1 \rangle + e^{i \gamma} \sin (\theta/2) \ |2\rangle,
\label{state}
\end{equation}
where we changed $\phi \mapsto \gamma$ to conform with the notation in~\cite{Rosi2017}.
By specializing Eqs.\,\eqref{AAverage} and \eqref{Apure} to this setting, we, respectively, have 
\begin{align*}
\!\!\frac{{\langle\hat{a}\rangle}_{|s\rangle}}{g} = \frac{r_1 + r_2}{2}  
+  \frac{r_1-r_2}{2} \cos \theta +  |r| \cos \left(\varphi_r + \gamma \right) \sin \theta,
\end{align*}
and
\begin{align} 
\frac{(\Delta \hat{a}^2)_{|s\rangle}}{g^2} &= \bigg(\frac{r_1 - r_2}{2}\! \bigg)^{\!2}\! \!\sin^2 \theta \notag \\
 &+  |r|^2 \Big[ 1- \cos^2(\varphi_r +\gamma) \sin^2 \theta \Big] \nonumber \\
& - \frac{|r|}{2}  \bigg(\frac{r_1-r_2}{2} \!\bigg)\cos(\varphi_r+\gamma) \sin(2 \theta)  .
\label{varas}
\end{align}

Let us now suppose that the angle $\theta$ differs from the optimal value $\theta=\pi/2$ by an amount $\epsilon \ll 1$. Then, we find 
\begin{equation*}
\!\!\frac{{\langle\hat{a}\rangle}_{|s\rangle}}{g} 
= \frac{r_1+r_2}{2} 
- \frac{r_2-r_1}{2} \sin \epsilon + |r| \cos(\varphi_r+\gamma) \cos\epsilon .
\end{equation*}
Notice that in the approximation $\sin \epsilon \approx \epsilon$, this yields a linear correction from the second term in the 
right-hand side. Due to its independence upon $\gamma$, this correction remains even after averaging over many realizations 
of the experiment. For the variance of the acceleration, using Eq.\,\eqref{varas} and averaging over $\gamma$ as in Eq.\,\eqref{AvPhi}, one finds
\begin{eqnarray*}
\frac{(\Delta \hat{a}^2)_{|s\rangle}}{g^2} 
\bigg\vert_{{\theta}= \frac{\pi}{2}+\epsilon} &\!\simeq\!&  \left(\frac{r_2-r}{2}\right)^2 + \frac{1}{2} |r|^2 \nonumber \\ 
& \! -\! & \bigg[\bigg(\frac{r_1-r_2}{2}\bigg)^2 - \frac{1}{2} |r|^2 \bigg] \frac{\epsilon^2}{2} +O(\epsilon^4), 
\end{eqnarray*}
showing that the variance is a more robust indicator than the mean acceleration with respect to small deviations
from the targeted superposition state, thanks to its quadratic dependence on $\epsilon$.

Furthermore,  the analysis in \cite{Rosi2017} yields a bound on the average value of the 
product $|r| \cos(\varphi_r+\gamma)$, rather than on $|r|$ itself. This is important because the average of 
$\cos(\varphi_r+\gamma)$ over $\gamma$ converges quickly to zero for $\gamma$ uniformly distributed in $[0, 2 \pi)$. 
This implies that, for a given bound on the average of $|r| \cos(\varphi_r+\gamma)$, increasingly larger values of $|r|$ are possible as the average of $\cos \gamma$ approaches zero. 
In order to give a quantitative idea of this effect, we plot in Fig.\,\ref{FigAppendix} (left) the distribution of values of $\cos \gamma$ for uniform random values of $\gamma$. In Fig.\,\ref{FigAppendix} (right), the ensemble average value of $\gamma$ is reported, together with its standard deviation, versus the 
size of the statistical ensemble. It is clear that $\gamma$ has a range of values compatible with a zero average even with a moderate statistics. The effect of its averaging becomes worse, from the standpoint of giving more stringent bounds to $|r|$, as the number of realizations is increased, defying the goal of a high-statistics sampling.

\begin{figure*}[t]
\includegraphics[width=8.5cm]{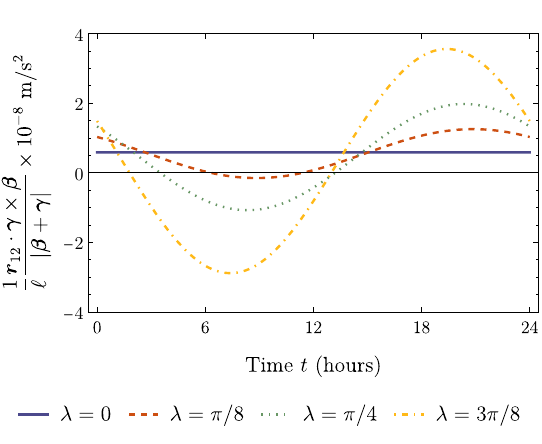}
\includegraphics[width=8.5cm]{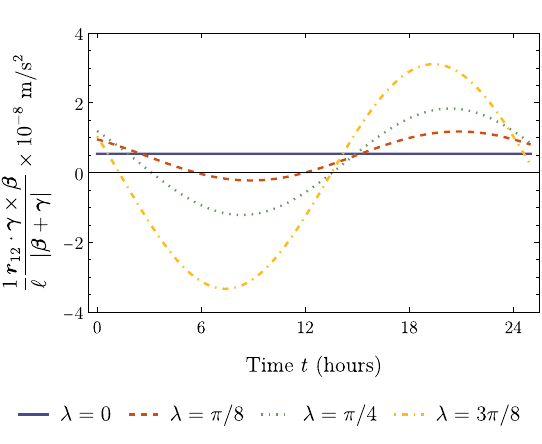}
\vspace*{-2mm}
\caption{Factor $\tau_0(t)$ in Eq.\,\eqref{torqueFactor} that appears in the definition of the torque operator $\hat{\tau}_\parallel$ [Eq.\,\eqref{torque2} in the main text], normalized by the balance length $\ell = | {\bf r}_{AB}|$, is plotted for various latitudes $\lambda$ of the torsion balance for (left) $\tilde{\theta} =0$ and (right) $\tilde{\theta} = \pi/4$. It is observed that this factor is relatively insensitive to the orientation of the balance $\theta$, relative to the plane spanned by ${\bf T} = -({\bf F}_A+ {\bf F}_B)$ and $\hat{{\bf z}}$.} 
\label{FigAppendixTorque}
\end{figure*}

\section{Computation of the vectors ${\boldsymbol{\beta}}(t)$ and ${\boldsymbol{\gamma}}(t)$}
\label{ComputationOfVectors}

With reference to the notation of Sec.\,\ref{sub:Torque} and the geometry depicted in Fig.\,\ref{Fig:eotvos}, we wish to express ${{\bf F}}_{M_\mathSun,j}(t)$ and ${{\bf F}}_{\bar{\Omega},j}$ in the lab frame. To this end, we first express the lab-frame unit vectors $\{\hat{{\bf x}},\hat{{\bf y}},\hat{{\bf z}}\}$ in the co-orbiting Earth-centered frame, with unit vectors $\{\hat{{{\bf \mathsf{x}}}},\hat{{{\bf \mathsf{y}}}},\hat{{{\bf \mathsf{z}}}}\}$:
\begin{eqnarray*}
\hat{{\bf x}} (t) &=& 
\begin{pmatrix}  
\cos \varepsilon & 0 & - \sin \varepsilon \\
0 & 1 & 0 \\
\sin \varepsilon & 0 & \cos \varepsilon \\
\end{pmatrix}  
\begin{pmatrix}  
\sin \lambda \cos \Omega t \\
 \sin \lambda \sin \Omega t \\
-\cos \lambda 
\end{pmatrix}  \nonumber \\
&=&
\begin{pmatrix}
\cos \varepsilon \sin \lambda \cos \Omega t + \sin \varepsilon \cos \lambda \\ 
\sin \lambda \sin \Omega t  \\
\sin \varepsilon \sin \lambda \cos \Omega t - \cos \varepsilon \cos \lambda \\ 
\end{pmatrix}_{\{ \hat{\mathsf{{\bf x}}}, \hat{\mathsf{{\bf y}}},\hat{\mathsf{{\bf z}}} \}},\\ 
 \hat{{\bf y}} (t) &= &
\begin{pmatrix}  
\cos \varepsilon & 0 & - \sin \varepsilon \\
0 & 1 & 0 \\
\sin \varepsilon & 0 & \cos \varepsilon \\
\end{pmatrix}  
\begin{pmatrix}  
\sin \Omega t \\
-\cos \Omega t \\ 0
\end{pmatrix}  \nonumber \\
&=&\begin{pmatrix}
\cos \varepsilon  \sin \Omega t \\ 
- \cos \Omega t  \\
\sin \varepsilon \sin \Omega t 
\end{pmatrix}_{\{ \hat{\mathsf{{\bf x}}}, \hat{\mathsf{{\bf y}}},\hat{\mathsf{{\bf z}}} \}}, 
\end{eqnarray*}
\begin{eqnarray*}
\hat{{\bf z}} (t) &=& 
\begin{pmatrix}  
\cos \varepsilon & 0 & - \sin \varepsilon \\
0 & 1 & 0 \\
\sin \varepsilon & 0 & \cos \varepsilon \\
\end{pmatrix}  
\begin{pmatrix}  
\cos \lambda \cos \Omega t \\
 \cos \lambda \sin \Omega t \\
\sin \lambda 
\end{pmatrix}  \nonumber \\
&=&\begin{pmatrix}
\cos \varepsilon \cos \lambda \cos \Omega t - \sin \varepsilon \sin \lambda \\ 
\cos \lambda \sin \Omega t  \\
\sin \varepsilon \cos \lambda \cos \Omega t + \cos \varepsilon \sin \lambda \\ 
\end{pmatrix}_{\{ \hat{\mathsf{{\bf x}}}, \hat{\mathsf{{\bf y}}},\hat{\mathsf{{\bf z}}} \}}, 
\end{eqnarray*}
where $\varepsilon \approx 23.4^\circ$ is the tilt of the Earth's rotation axis relative the ecliptic plane (see Fig.\,\ref{Fig:EarthSun}). Then, we may write 
\begin{eqnarray*}
\hat{{\bf d}}(t) &=& d_x(t) \hat{{\bf x}} + d_y(t) \hat{{\bf y}} + d_z(t) \hat{{\bf z}}, \nonumber \\ 
\hat{\bf R}_{M_\mathSun}(t) &=& R_{M_{\mathSun x}}(t) \hat{\bf x} + 
                                              R_{M_{\mathSun y}}(t) \hat{\bf y} + 
                                              R_{M_{\mathSun z}}(t) \hat{\bf z},
\end{eqnarray*}
where
\begin{eqnarray*}
d_x(t) =\hat{{\bf x}}(t) \cdot {\bf d} &\!=\!&\hat{{\bf x}}(t) \cdot  \left[{\bf R}_{\mathSun} - {\bf R}_\mathTerra(t)\right] \nonumber \\
 &\!=\!& \hat{{\bf x}}(t) \cdot \left[{R}_\mathSun  \hat{{\bf \mathsf{x}}} 
 - {R}_{\mathTerra} \hat{{\bf z}}(t)\right] \nonumber \\
 &\!=\!& {R}_{\mathSun} \left[ \sin \lambda  \cos \varepsilon  \cos \Omega t +\cos \lambda  \sin \varepsilon \right], \nonumber \\
    d_y(t) =\hat{{\bf y}}(t) \cdot {\bf d} &\!=\!& R_{\mathSun} \cos \varepsilon  \sin  \Omega t , \nonumber \\
    d_z(t) =\hat{{\bf z}}(t) \cdot {\bf d} &\!=\!& R_{\mathSun} \left[ \cos \lambda  \cos \varepsilon  \cos  \Omega t - \sin \lambda  \sin \varepsilon  \right]  \nonumber\\
    & \!-\!& R_{\mathTerra}, 
\end{eqnarray*}
and
\begin{eqnarray*}
    R_{\mathSun x}(t) &\!=\!&\hat{{\bf x}}(t) \cdot {\bf R}_\mathSun = R_\mathSun \hat{{\bf x}}(t) \cdot \hat{{\bf \mathsf{x}}}\nonumber \\ 
    &=& R_\mathSun \left[ \cos \varepsilon \sin \lambda \cos \Omega t + \sin \varepsilon \cos \lambda  \right], \nonumber \\
     R_{\mathSun y}(t) &\!=\!&\hat{{\bf y}}(t) \cdot {\bf R}_\mathSun = R_\mathSun \cos \varepsilon  \sin \Omega t,\nonumber \\
    R_{\mathSun z}(t) &\!=\!& \hat{{\bf z}}(t) \cdot {\bf R}_\mathSun = R_\mathSun \left[ \cos \varepsilon \cos \lambda \cos \Omega t - \sin \varepsilon \sin \lambda \right]. \nonumber
\end{eqnarray*}

Having explicitly computed $\hat{{\bf d}}(t)$ and ${\bf R}_\mathSun(t)$, we can now obtain ${\boldsymbol{\beta}}(t)$ and ${\boldsymbol{\gamma}}(t)$:
\begin{eqnarray}
\!\!\!\!{\boldsymbol{\beta}}(t) &\!\!=\!\!&\Omega^2 R_{\mathTerra} \cos \lambda \left( \sin \lambda \hat{{\bf x}}  +  \cos \lambda \hat{{\bf z}}\right) + \bar{\Omega}^2{{\bf R}}_\mathSun(t), 
\label{betaexpress} \\
\!\!\!\!{\boldsymbol{\gamma}}(t) &\!\!=\!\! &\frac{G_N M_{\mathSun}  }{d(t)^3} {{\bf d}}(t) - g_{\mathTerra} \hat{{\bf z}}.
\label{gammaexpress}
\end{eqnarray}
We parametrize the vector pointing from mass 1 to mass 2 in terms of the angles $\tilde{\theta}$ and $\tilde{\phi}$, 
\begin{equation*}
    {\bf r}_{AB} = \ell \begin{pmatrix}
        \cos \tilde{\phi} \cos \tilde{\theta} \\
        \cos \tilde{\phi} \sin \tilde{\theta} \\
        \sin \tilde{\phi}
    \end{pmatrix},
\end{equation*}
where $\ell = \|{\bf r}_{AB}\|$ is the length of the torsion balance, $\tilde{\theta}$ is a controllable angle corresponding to the orientation of the balance relative to the $x$ axis in the lab frame, and $\tilde{\phi}$ is the angle between the tension ${\bf T}=-({\bf F}_A+{\bf F}_B)$ and the $z$ axis, given by $\cos \tilde{\phi} = \hat{{\bf T}}\cdot \hat{{\bf z}}$.

Having determined the the vectors $\boldsymbol{\gamma}(t)$ and $\boldsymbol{\beta}(t)$, we can quantize the torque parallel to the fiber by  substituting the mass operators $\hat{M}_{i,j} = m_{i,j} \hat{I} + \hat{H}_{i,j}/c^2$ and $\hat{M}_{g,j} = m_{g,j} \hat{I} + \hat{H}_{g,j}/c^2$ into Eq.\,\eqref{torque}, which yields
\begin{eqnarray}
\hat{\tau}_\parallel(t) &\!=\!& \tilde{N}(t) \left[ \left(\frac{m_{g,A}}{m_{i,A}} - \frac{m_{g,B}}{m_{i,B}} \right) \left(  \hat{I} + \hat{T}_1 - \hat{T}_2(t) \right) + \hat{\Delta}\right] \nonumber \\
&& + \,\mathcal{O}\!\left(1/c^4\right),
\label{generalTorqueOperator}
\end{eqnarray}
 in terms of a time-dependent scalar factor 
\begin{align*}
 \tilde{N}(t)&=\frac{m_{i,A}m_{i,B} \left[  {\bf r}_{AB} \cdot  {\boldsymbol{\gamma}}(t) \times {\boldsymbol{\beta}}(t) \right]}{|  \left(m_{i,A} +m_{i,B}\right) {\boldsymbol{\beta}}(t)+ \left(m_{g,A} +m_{g,B}\right) {\boldsymbol{\gamma}}(t)|},
\end{align*}
 and operators 
 \begin{align*}
  \hat{T}_1 &= \frac{\hat{H}_{i,A}}{m_{i,A}c^2} + \frac{\hat{H}_{i,B}}{m_{i,B}c^2} , \nonumber \\
  \hat{T}_2(t) &= {{ \Big[ {| \boldsymbol{\beta}(t) \!
  \left(m_{i,A}+m_{i,B}\right) + \boldsymbol{\gamma }(t) \!\left(m_{g,A}+m_{g,B}\right) |}^2 c^2 \Big]}^{-1}}  
 \nonumber \\
 &\quad  \times \Big\{ \boldsymbol{\beta}(t)^2  \left( m_{i,A}+m_{i,B} \right) \big(\hat{H}_{i,A}+  \hat{H}_{i,B} \big)  \nonumber  \\
 & \quad +\boldsymbol{\gamma }(t)^2 \left(m_{g,1}+m_{g,2}\right) \big(\hat{H}_{g,A}+ \hat{H}_{g,B} \big) \nonumber \\
& \quad + 2 \boldsymbol{\beta} (t)\cdot {\boldsymbol{\gamma}}(t) 
 \Big[ \left(m_{i,A}+m_{i,B}\right) \big(\hat{H}_{g,A}  + \hat{H}_{g,A} \big) \nonumber \\ 
 & \quad +\left(m_{g,A}+m_{g,B}\right)\big(\hat{H}_{i,A}  + \hat{H}_{i,B} \big) \Big] \Big\}, \\
 \hat{\Delta} &= 
 \frac{ m_{i,A} \hat{H}_{g,A}  - m_{g,A} \hat{H}_{i,A}  }{m_{i,A}^2 c^2} \! - \! \frac{ m_{i,B} \hat{H}_{g,B}  - m_{g,B} \hat{H}_{i,B}  }{m_{i,B}^2 c^2}.
 \end{align*}

Under the assumption that $m_{g, j} =m_{i, j} =m_j$, Eq.\,\eqref{generalTorqueOperator} takes the simpler form given in the main text, Eq.\,\eqref{torque2},  In particular, we may now compute the factor $\tau_0(t)$ defined in Eq.\,\eqref{torqueFactor} in the main text, which 
involves the vectors ${\bf r}_{AB}$, ${\boldsymbol{\beta}}$, and ${\bf \tau}$ that appears in the torque 
operator $\hat{\tau}_\parallel$ in Eq.\,\eqref{torque2}.  A plot of $\tau_0(t)/\mu$ over the course of a 24-hour period is provided in 
Fig.\,\ref{FigAppendixTorque}. It is seen that this factor varies between approximately $ \pm 10^{-8} \,\text{m/s$^2$}$ throughout the course of a day.

\section{Computation of the angular acceleration in \\ the dynamical Cavendish experiment}
\label{CavendishAppendix}

With reference to the dynamical torsion-balance setting of Sec.\,\ref{Dynsetup}, we wish to determine the leading contributions to the angular acceleration operator due to the gravitational field. Substituting Eq.\,\eqref{dynamicalTorque} into the defining equation Eq.\,\eqref{eq:angacc}, and only retaining terms up to leading order in $1/c^2$, we find:

\onecolumngrid
\begin{eqnarray}
&&2 m R^2_t \,\hat{\alpha}(t)= \frac{ 2m \big[{\bf R}_t \times {\bf g}_A(t) \hat{M}_{g,A} - {\bf R}_t \times {\bf g}_B(t) \hat{M}_{g,B} \big]}{  \hat{M}_{i,A} +   \hat{M}_{i,B} }= \frac{{\bf R}_t \times {\bf g}_A(t) \hat{M}_{g,A} - {\bf R}_t \times {\bf g}_B(t) \hat{M}_{g,B}}{1 + \frac{\hat{H}_{i,A}/c^2 + \hat{H}_{i,B}/c^2}{2m}} \approx  \nonumber \\
&& \qquad\qquad \bigg( 1 - \frac{\hat{H}_{i,A}/c^2 + \hat{H}_{i,B}/c^2}{2m} \bigg) \left[{\bf R}_t \times {\bf g}_A(t) \left(m + \hat{H}_{g,A}/c^2\right) - {\bf R}_t \times {\bf g}_B(t) \left(m + \hat{H}_{g,B}/c^2\right)\right] \approx \nonumber \\
&&  m {\bf R}_t \times \left[ {\bf g}_A(t) - {\bf g}_B(t)\right] +  {\bf R}_t \times \left[ {\bf g}_A(t) \hat{H}_{g,A}/c^2 - {\bf g}_B(t) \hat{H}_{g,B}/c^2 \right]  -  m{\bf R}_t \times \left[ {\bf g}_A(t) - {\bf g}_B(t)\right] \frac{\hat{H}_{i,A}/c^2 + \hat{H}_{i,B}/c^2}{2m} 
\approx \nonumber \\
&& m {\bf R}_t \times \left[ {\bf g}_A(t) - {\bf g}_B(t)\right] +  \frac{1}{2 c^2}{\bf R}_t \times \left[ 2{\bf g}_A(t)\hat{H}_{g,A} - 2{\bf g}_B(t)\hat{H}_{g,B} \right.  -  \left. \left(  {\bf g}_A(t) - {\bf g}_B(t)\right) \left(\hat{H}_{i,A} + \hat{H}_{i,B} \right)\right] 
\nonumber = \\
&& m {\bf R}_t \times \left[ {\bf g}_A(t) - {\bf g}_B(t)\right] +  \frac{1}{2 c^2}{\bf R}_t \times \left[ {\bf g}_A(t) \left( 2\hat{H}_{g,A} - \hat{H}_{i,A} - \hat{H}_{i,B} \right)  -   {\bf g}_B(t) \left( 2\hat{H}_{g,B} - \hat{H}_{i,A} - \hat{H}_{i,B} \right)\right] .
\label{AngularAcceleration}
\end{eqnarray}
Given the geometry depicted in Fig.\,\ref{fig:Cavendish} and the gravitational fields at $A$ and $B$ as in Eqs.\,\eqref{gA}--\eqref{gB}, and using the definition of $R_\pm(t)$ in Eq.\,\eqref{Carnot}, the cross products appearing in $\hat{\alpha}_g(t)$ simplify to
\begin{eqnarray*}
    {\bf r} \times {\bf g}_A(t) &&= +\frac{G_N m_s}{{R}_{-}^3} {\bf R}_t \times \left( {\bf R}_t - {\bf R}_s  \right) + \frac{G_N m_s}{{R}_{+}^3}{\bf R}_t 
    \times \left( {\bf R}_t + {\bf R}_s  \right)=  \nonumber \\
    &&= -\frac{G_N m_s}{{R}_{-}^3} {\bf R}_t \times {\bf R}_s + \frac{G_N m_s}{{R}_{+}^3} {\bf R}_t \times {\bf R}_s  
    = G_N m_s \left(\frac{1}{{R}_{+}^3} -\frac{1}{{R}_{-}^3} \right) {\bf R}_t \times {\bf R}_s, \\
    {\bf r} \times {\bf g}_B(t) &=&+\frac{G_N m_s}{{R}_{+}^3} {\bf r} \times \left( - {\bf R}_t - {\bf R}_s  \right) +  
    \frac{G_N m_s}{{R}_{-}^3} {\bf R}_t \times \left( -{\bf R}_t + {\bf R}_s  \right) = \nonumber \\ 
    &&-\frac{G _N m_s}{{R}_{+}^3} {\bf r} \times{\bf R}_s +  \frac{G_N m_s}{{R}_{-}^3} {\bf R}_t \times {\bf R}_s  
    =- G_N m_s \left(\frac{1}{{R}_{+}^3} -\frac{1}{{R}_{-}^3} \right) {\bf R}_t \times {\bf R}_s.
\end{eqnarray*}
Upon substituting the above results into Eq.\,\eqref{AngularAcceleration}, the angular acceleration operator becomes
\begin{eqnarray*}
\hat{\alpha}(t) &=& \frac{G_N m_s}{2 R_t^2}  \left(\frac{1}{{R}_{+}^3} -\frac{1}{{R}_{-}^3} \right) {\bf R}_t \times {\bf R}_s 
\left[ 2  + \frac{1}{mc^2} \left(  \hat{H}_{g,A} - \hat{H}_{i,A} + \hat{H}_{g,B} - \hat{H}_{i,B}  \right) \right] \nonumber \\
&=& \frac{G_N m_s}{R_t^2}
 \left(\frac{1}{{R}_{+}^3} -\frac{1}{{R}_{-}^3} \right) R_t  R_s \sin \left( \Theta -\Omega t\right) \left[ 1  + \frac{1}{2mc^2} \left(  \hat{H}_{g,A} - \hat{H}_{i,A} + \hat{H}_{g,B} - \hat{H}_{i,B}  \right) \right] \hat{{\bf z}}\nonumber \\
&=& G_N m_s \frac{R_s}{R_t}  \left(\frac{1}{{R}_{+}^3} -\frac{1}{{R}_{-}^3} \right)  \sin  \left( \Theta -\Omega t\right) 
  \left[ 1  + \frac{1}{2} \left( \frac{ \hat{H}_{g,A} - \hat{H}_{i,A}}{m c^2} + \frac{\hat{H}_{g,B} - \hat{H}_{i,B}}{m c^2}  \right) \right] \hat{{\bf z}} .
\end{eqnarray*}
Using Eq.\,\eqref{HamiltonianAndr} to reexpress the differences in the last term for $j=A,B$ yields Eq.\,\eqref{angularAcceleration} in the main text.

\twocolumngrid

\end{document}